\documentclass[10pt,journal,compsoc]{IEEEtran}
\ifCLASSOPTIONcompsoc
  \usepackage[nocompress]{cite}
\else
  \usepackage{cite}
\fi
\ifCLASSINFOpdf
  \usepackage[pdftex]{graphicx}
  \graphicspath{{./Figures/}}
  \DeclareGraphicsExtensions{.pdf,.jpg,.jpeg,.png,}
\else
\fi
\usepackage{graphicx}
\usepackage{amsmath,amssymb} 
\usepackage{color}
\usepackage{multirow}
\usepackage[flushleft]{threeparttable} 
\usepackage{boldline} 
\usepackage[table]{xcolor} 
\usepackage[colorlinks=true,allcolors=black]{hyperref}

\usepackage{dcolumn}
\usepackage{bm}

\usepackage[utf8]{inputenc}
\usepackage[T1]{fontenc}
\usepackage{mathptmx}

\usepackage{graphicx,amsfonts,amsmath,amssymb,xcolor,amsthm}
\usepackage{graphicx,amsfonts,amsmath,amssymb,xcolor,amsthm}

\usepackage{algorithm}
\usepackage[noend]{algpseudocode}

\newcommand{\argmin}{\arg\!\min}

\newcommand{\Dmat}{{\bf D}}
\newcommand{\Emat}[0]{{{\bf E}}}

\newcommand{\Imat}{{\bf I}}

\newcommand{\Mmat}[0]{{{\bf M}}}

\newcommand{\Tmat}[0]{{{\bf T}}}

\newcommand{\Xmat}{{\bf X}}
\newcommand{\Ymat}[0]{{{\bf Y}}}
\newcommand{\Zmat}{{\bf Z}}

\newcommand{\cv}{{\boldsymbol{c}}}

\newcommand{\ev}[0]{{\boldsymbol{e}}}

\newcommand{\pv}[0]{{\boldsymbol{p}}}

\newcommand{\uv}[0]{{\boldsymbol{u}}}
\newcommand{\vv}{\boldsymbol{v}}

\newcommand{\xv}{\boldsymbol{x}}
\newcommand{\yv}{\boldsymbol{y}}

\newcommand{\zv}{\boldsymbol{z}}

\newcommand{\Thetamat}{\boldsymbol{\Theta}}

\newcommand{\Phimat}{\boldsymbol{\Phi}}

\newcommand{\ts}{^{\top}}
\newcommand{\ie}{{\em i.e.}}

\newcommand{\etc}{{\em etc.}}

\newcommand{\inv}{^{-1}}

\begin{document}
%
\title{Snapshot Interferometric 3D Imaging by Compressive Sensing and Deep Learning}
%
%
%

\author{Mu Qiao, Yangyang Sun, Jiawei Ma, Ziyi Meng, Xuan Liu,~\IEEEmembership{Member,~IEEE} and
        Xin~Yuan,~\IEEEmembership{Senior~Member,~IEEE}
\IEEEcompsocitemizethanks{
\IEEEcompsocthanksitem Q.~Mu and X. Liu are with Department of ECE, New Jersey Institute of Technology, Newark, NJ 07102, USA. E-mail: \{mu.qiao, xuan.liu\}@njit.edu. \protect
\IEEEcompsocthanksitem Y. Sun is with College of Optics and Photonics (CREOL), University of Central Florida, Orlando, FL 3281, USA. E-mail: yangyang@knights.ucf.edu. \protect
\IEEEcompsocthanksitem J. Ma is with Department of Computer Science, Columbia University, New York, NY 10027, USA. E-mail: jiawei.m@columbia.edu. \protect
\IEEEcompsocthanksitem Z. Meng is with State Key Laboratory of Information Photonics and Optical Communications, Beijing University of Posts and Telecommunications, Beijing, 100876, China and also with Department of ECE, New Jersey Institute of Technology, Newark, NJ 07102, USA. E-mail: zm233@njit.edu.
\IEEEcompsocthanksitem X.~Yuan is with Bell Labs, Murray Hill, New Jersey, 07974 USA.\protect\\
E-mail: xyuan@bell-labs.com (corresponding author).
}
\thanks{Manuscript updated \today.}
}

\IEEEtitleabstractindextext{%
\begin{abstract}
We demonstrate single-shot compressive three-dimensional (3D) $(x, y, z)$ imaging based on interference coding. The depth dimension of the object is encoded into the interferometric spectra of the light field, resulting a $(x, y, \lambda)$ datacube which is subsequently measured by a single-shot spectrometer. 
By implementing a compression ratio up to $400$, we are able to reconstruct $1G$ voxels from a 2D measurement.
Both an optimization based compressive sensing algorithm  and a deep learning network are developed for 3D reconstruction from a single 2D coded measurement. Due to the fast acquisition speed, our approach is able to capture volumetric activities at native camera frame rates, enabling 4D (volumetric-temporal) visualization of dynamic scenes.
\end{abstract}

}

\maketitle

\section{Introduction\label{sec:introduction}}
In optical imaging, zero dimensional data can be directly acquired using zero dimensional sensor, such as a photodiode. Similarly, one dimensional and two dimensional data can be acquired using one dimensional sensors (line-scan CCD or CMOS cameras) or two dimensional sensors (area-scan CCD or CMOS cameras).
Currently, sensors that directly perform three-dimensional (3D) data acquisition do not exist.
Hence, data acquisition for volumetric optical imaging remains challenging. In this paper, we describe a novel snapshot interferometric 3D imaging (SI3D) technology that takes 2D compressive measurement for 3D imaging. The SI3D system encodes the depth information of the 3D object in interferometric spectrum and modulates the interferometric signal originating from a specific lateral coordinate with a unique random pattern. After this, interferometric signals originating from different lateral coordinates are superimposed and detected by a 2D sensor (2D camera). We then recover interferometric spectra from different lateral coordinates by pursuing (or learning) data structure (with sparsity as a special case), and reconstruct the 3D object by performing Fourier analysis on individual spectra. By taking an innovative compressive sensing approach, SI3D enables accurate reconstruction of interferometric signals from different lateral coordinates from the measurement of their superimposition. SI3D takes a single-shot approach for data acquisition and achieves an extremely large bandwidth for 3D data acquisition, using a 2D camera that has a moderate frame rate.

Snapshot compressive imaging (SCI) systems aim to capture high dimensional ($\ge 3$) data with a single-frame of 2D measurement in a multiplexing fashion, which eliminates any scanning effort and enables high acquisition rates  (thus high throughput) and low bandwidth/memory requirement. Equipped with advanced compressive sensing (CS)~\cite{Donoho06ITT,Candes06ITT} reconstruction algorithms, SCI has demonstrated promising results on $x$-$y$-spectral~\cite{Gehm07_DD,wagadarikar2008single,Arce14_SPM_CASSI,Cao16_SPM_CASSI,Lin14_ACMTG,Cao2011_Cao_Prism}, -polarization~\cite{Tsai15OE}, -temporal~\cite{gao2014single,qi2020single,llull2013coded,Qiao2020_CACTI}, -spectral-temporal~\cite{Tsai15OL} and -depth~\cite{llull2015image,antipa2018diffusercam} data.
Inspired by recent advances in deep learning (DL)~\cite{Goodfellow-et-al-2016}, convolutional neural networks (CNNs) have also been adapted to these SCI systems~\cite{Barbastathis19DL,Qiao2020_APLP,Yuan2020_CVPR_PnP} for faster and more accurate reconstruction.

In this paper, we describe a SCI 3D $(x, y, z)$ imaging approach that captures a single 2D compressed measurement for 3D reconstruction. Our approach first encodes the depth information of the 3D object into spectral dimension of the light field through an interferometer, resulting in a $(x, y, \lambda)$ datacube that is subsequently sampled by an single-disperser coded-aperture spectrometer~\cite{wagadarikar2008single}.
We then recover the $(x, y, \lambda)$ datacube through CS or DL algorithms and finally reconstruct the 3D object by performing an inverse conversion from the spectral to the depth dimension.

The interference-based depth encoding mechanism of our approach takes the following advantages over previous diffraction-based depth encoding strategies~\cite{brady2009compressive,antipa2018diffusercam}. 
First, a higher and uniform axial resolution (up to $13\mu m$) is achieved across the entire depth field of view (up to $1.6mm$). 
Second, the conversion of the datacube from the $(x, y, z)$ space to $(x, y, \lambda)$ space allows us to design an image-space coding setup~\cite{wagadarikar2008single,gao2014single,llull2013coded} as opposed to the pupil coding scheme~\cite{asif2015flatcam,stork2014optical,ozcan2016lensless,harm2014lensless}, which dramatically reduces the computation complexity of the reconstruction and enables us to reconstruct $1G$ voxels from a 2D measurement of $5.5M$ ($2560\times2160$) pixels with a med-end computer. 
Further, the image-space coding scheme requires much less calibration effort, because only a single image of the coded aperture is required for the reconstruction. The simplified calibration process also leads to improved robustness and stability.
Moreover, a deep learning network is developed to perform the large-scale inversion in our system, which reduces the reconstruction time from hours to sub-seconds.

The rest of this paper is organized as follows. Section~\ref{Sec. related work} compares our approach with some related works. In Section~\ref{Sec. method}, we describe the principle of our approach in detail including the depth-encoding mechanism and image reconstruction through both CS and deep learning algorithms. In Section~\ref{Sec. system} and~\ref{Sec:3D_result}, we describe hardware implementation and show several 3D dynamic scenes captured by our system. Section~\ref{Sec:con} concludes the paper. System performance is characterized in Appendix~\ref{Sec:performance}.

\section{Related work}
\label{Sec. related work}

In optical imaging, data with dimensionality $\le2$ can be directly measured using sensors of matched dimensionality.
However, sensors that directly perform three-dimensional (3D) data acquisition do not exist.
Therefore, scanning is often required in 3D imaging such as confocal ~\cite{pawley2010handbook}, two-photon~\cite{denk1990two}, light-sheet microscopy~\cite{chen2014lattice}, and optical coherence tomography~\cite{huang1991optical}. Alternatively, advanced algorithms can be applied to data acquired from multiple 2D measurements to recover the 3D image, such as in computed tomography~\cite{brenner2007computed}, diffraction tomography~\cite{lauer2002new}, structured illumination~\cite{schermelleh2008subdiffraction}, stereo~\cite{brown2003advances}, depth from focus and defocus~\cite{xiong1993depth} techniques.
However, these techniques often suffer from low speed due to the sequential acquisition fashion.
To improve the speed, some single-shot techniques have been developed which distribute 3D information onto a 2D measurement. For examples, plenoptic cameras (a.k.a. light field cameras)~\cite{levoy2006light} deploy 2D microlens arrays before 2D detectors to simultaneously capture 2D intensity distributions and 2D light direction distributions, from which depth can be inferred.
However, this class of single-shot techniques usually has low spatial resolutions because the limited 2D pixels have to be distributed in higher-dimensional space.

Compressive sensing theory~\cite{Donoho06ITT,Candes06ITT} provides an elegant solution to overcome this speed-resolution trade-off by exploiting structural features in the dataset. The CS theory states that a signal can be fully reconstructed with much less measurements than what is prescribed by Nyquist sampling rate, given the signal is sparse in some basis. 
Several CS-based 3D imaging approaches have been demonstrated by encoding the depth information either into the speckle response of a scattering medium~\cite{antipa2018diffusercam} or a complex hologram~\cite{brady2009compressive,zhang2018twin,choi2010compressive}. Some prior works have also leveraged CS in diffraction tomography~\cite{brady2015compressive,li2009compressive} for single-shot measurement. All these techniques encode depth through light diffraction and therefore give a depth resolution equal to the axial isoplanatic patch~\cite{brady2009compressive}, which is given by $\delta_z =\lambda (2z)^2/D^2$ with $\lambda$ being the wavelength, $D$ being the size of the effective aperture and $z$ the distance between the object and the effective aperture plane. $\delta_z$ is usually on the scale of sub-millimeter or larger and  generally depends on imaging depth. In contrast, our approach employs {\em interference} for more sensitivity depth sensing and achieve a uniform depth resolution (tens of micrometers) at least one order of magnitude higher than those based-on light diffraction.


Single-shot CS systems typically have two regimes in term of coding strategy: pupil coding and image-space coding. 
The former is usually adopted in lensless cameras where a random medium, be it a scattering medium~\cite{antipa2018diffusercam,liutkus2014imaging}, a mask~\cite{stork2014optical} or a random reflective surface~\cite{stylianou2016sparklegeometry}, is employed in lieu of a lens in front of the 2D detector. 
Each point in the object space produces a distinguishing random pattern, rather than a spot, on the 2D detector. To reconstruct the object, a 2D sensing matrix is required which is constructed by vectorizing the random patterns corresponding to each location in the object space as each column of the matrix. Therefore, the computation complexity, determined by the matrix size, scales with the square of the height of the measurement image (assuming a square image). 
By contrast, in the image-space coding regime, two cascaded relay lenses are typically employed - the first lens reproduces the object on the coded aperture plane and the second lens relays the coded object to the detector plane. At the same time, a modulator is placed on the coded aperture plane~\cite{llull2013coded}, or between the aperture and the detector~\cite{wagadarikar2008single}, or on the detector plane~\cite{gao2014single}, to shear the coded object cube before it collapses to a 2D measurement on the detector. In this way, the sensing matrix in the reconstruction does not necessarily have to be an expanded 2D matrix. Instead, it can be a 3D matrix with each page (each transverse plane) being a shifted version of the coded aperture pattern. Therefore, the matrix size scales only linearly with the height of the measurement image, dramatically reducing the computation complexity for large scale data. 
Some previous works~\cite{antipa2018diffusercam,brady2009compressive,zhang2018twin,choi2010compressive} have applied the pupil coding regime directly in the real 3D (volumetric) space. 
By contrast, our approach first converts the datastream from the 3D $(x, y, z)$ object into a spatial-spectral datacube $(x, y, \lambda)$, and then samples the $(x, y, \lambda)$ datacube using the image-space coding regime for more efficient reconstruction. This allows us to achieve a large scale reconstruction - $1G$ voxels recovered from a $5.5M$ pixels 2D measurement - with a med-end computer ($12$ CPU cores and $64$G memory). 
The image-space coding design also brings another important benefit of convenient calibration. Only a single image of the coded aperture needs to be captured for constructing the 3D sensing matrix, as different pages of the 3D  matrix are just shifted versions of the same aperture image. Some previous works also adopted image-space coding for 3D imaging, such as in~\cite{lin2013coded} and~\cite{llull2014compressive}. However, implementation of image-space coding directly in the real 3D space requires scanning the focal plane during the exposure, which limits the frame rates and increases system complexity.

Some techniques also utilize interference for depth sensing. For example, phase unwrapping~\cite{ghiglia1998two,wang2019one} methods infer depth from a measured phase map by assuming depth continuity across the boundaries between 0 and $2\pi$ in the phase map. Typically, sub-micrometer resolution can be achieved with an axial field of view (FoV) of several hundred micrometers~\cite{saldner1997profilometry} in these techniques. However, the reconstruction precision tends to be affected by steep depth change or dark area in the FoV which would break down the continuity assumption. 
Compressive holography~\cite{brady2009compressive,zhang2018twin,choi2010compressive} also uses interference to acquire holograms from which 3D can be reconstructed. However, as mentioned before, the depth resolution appears to be low (several sub-millimeters with an axial FoV of several centimeters) due to the large axial isoplanatic patch. Compared to these methods, our method exploits {\em interference of broadband light} rather than monochromatic light, and serves as a midpoint by providing a depth resolution of {\em a few tens of micrometers with a FoV of several millimeters}.


\begin{figure*}[t]
	\centering
	\fbox{\includegraphics[width=1\linewidth]{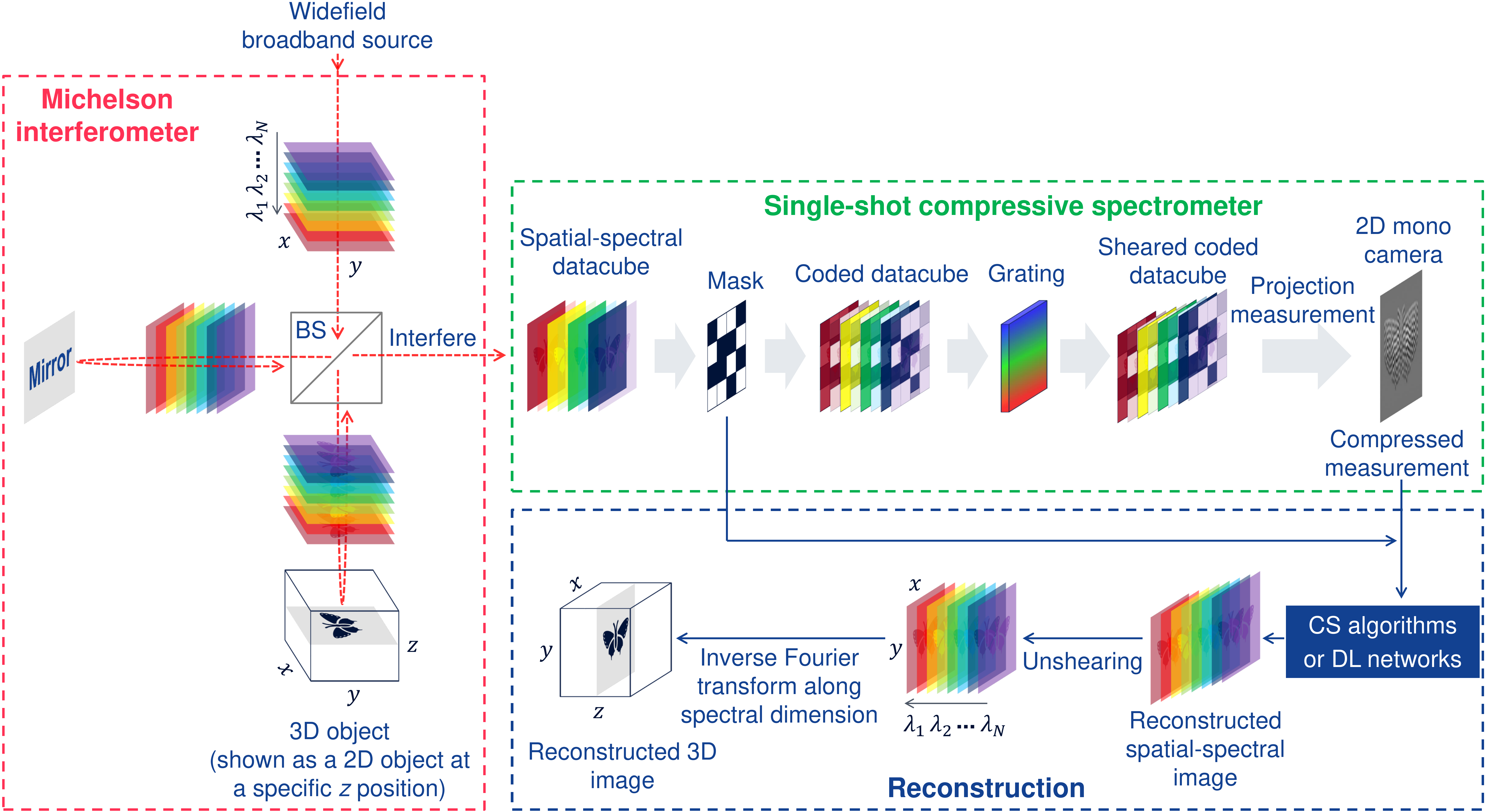}}
	\caption{Schematic of our approach. Our system consists of two parts: a wide-field broadband interferometer (left red box) and a single-shot compressive spectrometer (top-right green box). The interferometer converts the 3D information $(x,y,z)$ of the object into the spatial-spectral dimension $(x,y,\lambda)$ of the light field, which is then sampled by the spectrometer composed of a coded aperture (mask), a dispersive element (grating) and a 2D array detector (camera). The spatial-spectral datacube is first laterally modulated by the 2D mask in a manner of element-wise product for all the spectral channels. Then the modulated datacube is sheared by the grating along the dispersion direction (vertical in the diagram), resulting in an effectively coded datacube with each spectral channel coded by a distinguishing pattern (shifted version of the mask). This sheared datacube then collapses (integrated) along the spectral dimension into a 2D compressed image on a camera, which is then used in conjunction with the mask pattern to reconstruct the 3D image of the object through compressive sensing algorithms or deep learning networks (bottom-right blue box).}
	\label{fig:principle}
\end{figure*}

CS was originally proposed for sampling sparse signals and early inversion algorithms therefore aimed to search for the best sparse basis for high-fidelity reconstruction. Recently, inspired by the extensive advances in deep learning, researchers have been developing deep neural networks for CS inversion~\cite{Chang17ICCV,Iliadis18DSPvideoCS,Jin17TIP,Kulkarni2016CVPR,LearningInvert2017,Ma19ICCV,George17lensless,Barbastathis19DL,Yuan18OE,Miao19ICCV}.
DL provides some substantial advantages over the conventional optimization-based algorithms. On one hand, DL does not need to impose explicit priors to the unknown signal; instead, it {\em learns} the structure of the signal, which does not necessarily have to be sparsity, from massive training data through a deep convolutional neural network, which mitigates the restriction on signal types.
On the other hand, by constructing an end-2-end network, DL can directly output the desired reconstruction from an input measurement, in an almost instantaneous fashion, which significantly speeds up the reconstruction phase from hours to subseconds in CS imaging~\cite{Qiao2020_APLP}. 
In this paper, apart from a specially designed optimization-based algorithm, we also develop a DL algorithm for subseconds reconstruction for the proposed 3D imaging approach.

\section{Method}
\label{Sec. method}

\subsection{Depth encoding}
Fig.~\ref{fig:principle} illustrates the principle of our approach. We first convert the 3D information $(x, y, z)$ of the object into a spatial-spectral datacube $(x, y, \lambda)$ via a broadband Michelson interferometer. This conversion adopt the idea from optical coherence tomography~\cite{huang1991optical}. For simplicity, we consider the interference signal between the reference beam and the object beam reflected from a single axial line at $(x, y)$. The interference intensity $I$ for wavelength $\lambda$ can be expressed as:
\begin{equation}
I(\lambda)=\int I_r+I_s(z)+2\sqrt{I_r I_s (z)}\cos\left(2\pi\frac{2z}{\lambda}\right)dz,
\label{Eq:interference1}
\end{equation}
where $I_r$ is the intensity of the reference beam, $I_s(z)$ represents the intensity of the reflected light from object point $(x, y, z)$, and the factor $2$ in front of $z$ in the phase delay term originates from the round trip of the reflected light. Note that the reference mirror is assumed to be at $z=0$. Under uniform and unitary illumination, $I_s(z)$ represents the reflectivity of the object. By subtracting the direct current (DC) terms, we obtain the desired interference signal $I_{AC}$: 
\begin{equation}
I_{AC}(k)\propto\int\sqrt{I_s(z) }(e^{i2zk}+e^{-i2zk})dz,   
\label{Eq:interference2}
\end{equation}
where $k=\frac{2\pi}{\lambda}$ is the wavenumber. Form Eq.\eqref{Eq:interference2} we know that $z$ and $k$ are a Fourier transform pair. By performing inverse Fourier transform to $I_{AC}(k)$, we obtain $\sqrt{I_s(z)}+\sqrt{I_s(-z)}$, which is symmetric about the origin point. By placing the object on the positive part of the $z$-axis (\ie, placing the object above the reference mirror), we can simply discard the negative part and obtain $\sqrt{I_s(z)}$. In general, we can obtain the 3D distribution of the object $\sqrt{I_s(x,y,z)}$ by performing an inverse Fourier transform along the spectral dimension of the acquired interference signal - the spatial-spectral datacube $I_{AC}(x,y,k)$. Since $k$ and $\lambda$ are linearly connected if the bandwidth is relatively small, $z$ and $\lambda$ are also a Fourier transform pair. Therefore, the depth resolution $\delta_z$ is inversely proportional to the spectral width $\Delta_\lambda$ of the source (assume a Gaussian-shaped spectrum~\cite{izatt2015theory}):
\begin{equation}
\delta_z = 0.44 \frac{\lambda_0^2}{\Delta_\lambda},
\label{Eq:depth resolution}
\end{equation}
where $\lambda_0$ is the center wavelength. In analogy, the axial FoV $\Delta_z$ is inversely proportional to the spectral sampling interval $\delta_\lambda$:
\begin{equation}
\Delta_z=\frac{\lambda_0^2}{4\delta_\lambda}.
\label{Eq:depth fov}
\end{equation}

\subsection{CS sampling \label{Sec:Algo}}

We sample the spatial-spectral datacube $(x,y,\lambda)$ generated by the interferometer with a single-shot compressive spectrometer~\cite{wagadarikar2008single}. To implement effective compression, each spectral channel of the $(x,y,\lambda)$ datacube, \ie, each $(x,y)$ plane of different $\lambda$, needs to be coded/modulated with a distinguishing 2D random pattern before the datacube collapses along the spectral dimension to form the compressed measurement. In practice, applying different physical coded apertures to different spectral channel is hard to implement. Fortunately, however, by simply using a single coded aperture (mask) in conjunction with a dispersive element (grating), one can achieve the same effect. The mask first imposes the same random spatial modulation to all the spectral channels. Afterwards, the grating transversely shifts different spectral channels by different amount, resulting in a sheared datacube with each spectral channel coded by a shifted version of the mask pattern. The sheared datacube is then projected along the spectral dimension onto a 2D detector to form the compressed image (a.k.a., the measurement). From this compressed measurement, we can reconstruct the sheared datacube using compressive sensing algorithms or deep learning networks. The reconstructed datacube is subsequently unsheared and inverse Fourier transformed along the spectral dimension to finally recover the 3D image of the object.

Below we mathematically model the sampling process of our approach in the discrete form.
We start from the output of the interferometer.
Let $\Xmat \in {\mathbb R}^{N_x \times N_y \times N_{\lambda}}$ denote the spatial-spectral datacube with $N_\lambda$ spectral channels and $N_x\times N_y$ pixels in the transverse dimension. This spatial-spectral datacube is modulated by the mask $\Mmat^*\in {\mathbb R}^{N_x \times N_y}$ into $\Zmat \in {\mathbb R}^{N_x\times N_y \times N_\lambda}$. Specifically,
for each spectral channel, 
\begin{equation}
\Zmat_k=\Xmat_k\odot \Mmat^*,\quad\forall k=1,\cdots,N_\lambda,
\end{equation}
with $\odot$ denoting element-wise product. 
This modulated data cube $\Zmat$ is then sheared by the grating along the $y$ dimension into $\Zmat'\in {\mathbb R}^{N_x\times(N_y +N_\lambda-1) \times N_\lambda}$ 
with each spectral channel of $\Zmat'$ being 
\begin{equation}
\Zmat'_k (i,j)=\Zmat_k (i,j+g(\lambda_k-\lambda_c)),
\end{equation}
where $\lambda_k$ is the wavelength of $k^{th}$ channel, $\lambda_c$ denotes the center-wavelength of the broadband source and $g$ signifies the dispersion coefficient of the grating. 
This sheared coded cube is then summed along the spectral dimension to form the 2D compressed image on the camera, which can be represented as
\begin{equation}
    \Ymat = \sum_{k=1}^{N_{\lambda}} \Zmat'_k + \Emat,
\end{equation}
where $\Emat\in {\mathbb R}^{N_x\times(N_y +N_\lambda-1) \times N_\lambda }$  denotes the measurement noise. 
This equation can also be written as 
\begin{equation}
    \Ymat = \sum_{k=1}^{N_{\lambda}}\Xmat'_k \odot \Mmat_k + \Emat,
    \label{Eq: forward model 3D}
\end{equation}
where $\Xmat' \in {\mathbb R}^{N_x\times(N_y +N_\lambda-1) \times N_\lambda }$ is the sheared version of $\Xmat$, \ie,  
$\Xmat'_k (i,j)=\Xmat_k (i,j+g(\lambda_k-\lambda_c ))$, and $ \Mmat\in {\mathbb R}^{N_x\times(N_y +N_\lambda-1) \times N_\lambda }$ is a 3D modulation pattern imposed on $\Xmat'$ with $\Mmat_k (i,j)=\Mmat^*(i,j+g(\lambda_k-\lambda_c ))$.

\subsection{Optimization based reconstruction \label{Sec:opt}}

To implement the optimization based CS algorithm for image reconstruction, a 2D sensing matrix $\Phimat \in {\mathbb R}^{N_x (N_y+N_\lambda-1) \times N_x (N_y+N_\lambda-1) N_\lambda}$ associated with $\Mmat$ needs to be constructed by vectorizing $\Ymat$, $\Xmat'$ and $\Emat$ in ~\eqref{Eq: forward model 3D}. Let $\yv={\rm{vec}}(\Ymat)\in {\mathbb R}^{N_x (N_y+N_\lambda-1) }$, $\xv={\rm vec}(\Xmat')\in {\mathbb R}^{N_x (N_y+N_\lambda-1) N_\lambda}$ and $\ev={\rm vec}(\Emat)\in {\mathbb R}^{N_x (N_y+N_\lambda-1) }$; we have 
 \begin{equation}
 \Phimat =[\Dmat_1,\Dmat_2,\cdots,\Dmat_{N_\lambda}], \label{eq:Phimat} 
 \end{equation}
 where $\Dmat_k={\rm Diag}({\rm vec}(\Mmat_k ))$ denotes a diagonal matrix that has its diagonal entries filled by the vectorized form of $\Mmat_k$. We then arrive at
 \begin{equation}
   \yv=\Phimat \xv+ \ev. \label{Eq:yPhix} 
 \end{equation}
Eq.~\eqref{Eq:yPhix} shares the similar formulation as in the CS theory~\cite{Donoho06ITT,Candes06ITT}, which is an ill-posed equation. The theoretical analysis has recently been addressed in~\cite{Jalali19TIT_SCI} considering the special structure of the sensing matrix. 
To solve the problem, a regularization term $R(\xv)$ (a.k.a. priors) is usually required to confine the solution:
\begin{equation}
   \hat{\xv}= \argmin_{\xv} \|\yv- \Phimat\xv\|_2^2+\tau R(\xv),    \label{Eq:constrain}    
\end{equation}                                            
where $\|~\|_2$ denotes the $\ell_2$-norm and $\tau$ is a parameter to balance the fidelity term and priors. In this paper, we jointly use total variation~\cite{rudin1992nonlinear} and sparsity in wavelet domain~\cite{duarte2008wavelet} to account for piece-wise smooth and sparsity priors, respectively. We solve this problem by the alternating minimization framework~\cite{boyd2011distributed}.

After $\hat{\xv}$  is solved, we un-vectorize it to get $\Xmat'$ and unshear $\Xmat'$ to obtain $\Xmat$. By performing an inverse Fourier transform along the spectral dimension of $\Xmat$, we obtain the desired 3D image of the object.
Note that due to the special structure of $\Phimat$ in ~\eqref{eq:Phimat}, we never save this big 2D sensing matrix $\Phimat$ during the inversion, but use the 3D matrix $\{\Dmat_k\}_{k=1}^{N_{\lambda}}$ instead, which saves the memory as well as the running time. 
Algorithm details can be found in Appendix~\ref{Sec:Algorithm details}. 

\begin{figure*}[t]
	\centering
	\includegraphics[width=1\linewidth]{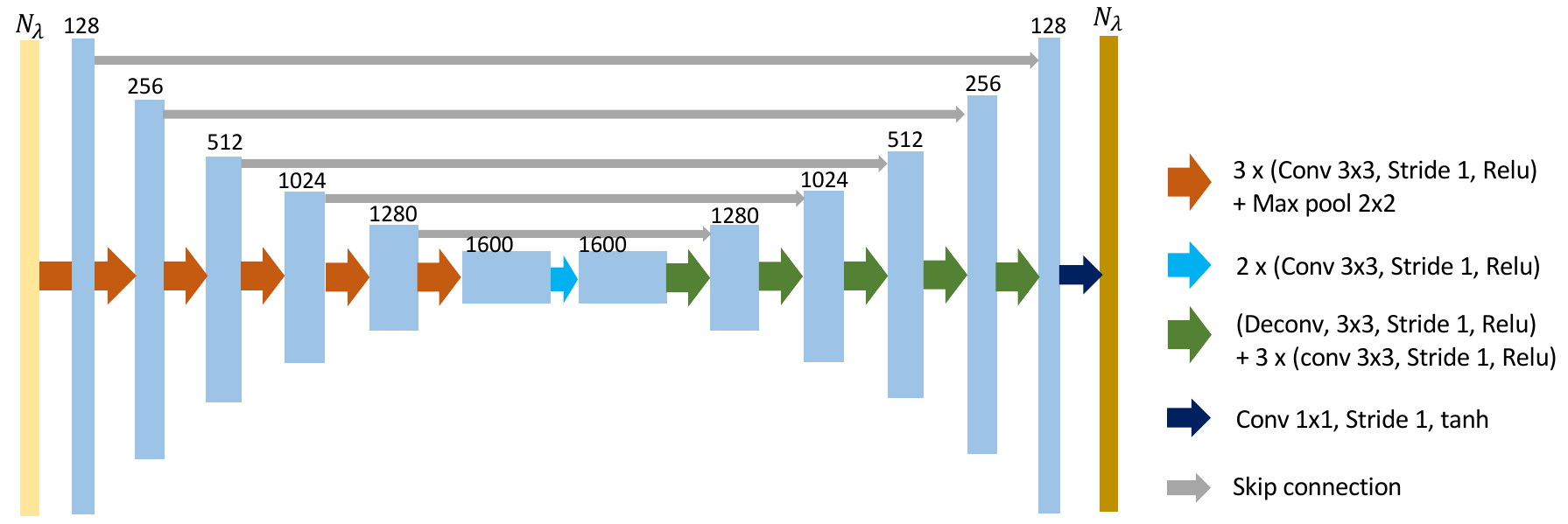}
	\caption{\label{fig:Unet} Diagram of CNN architecture for 3D reconstruction. $N_{\lambda}$ denotes the number of reconstructed spectral channels from one compressed measurement.}
\end{figure*}

\begin{figure}[htbp!]
    \centering
    \includegraphics[width=.9\linewidth]{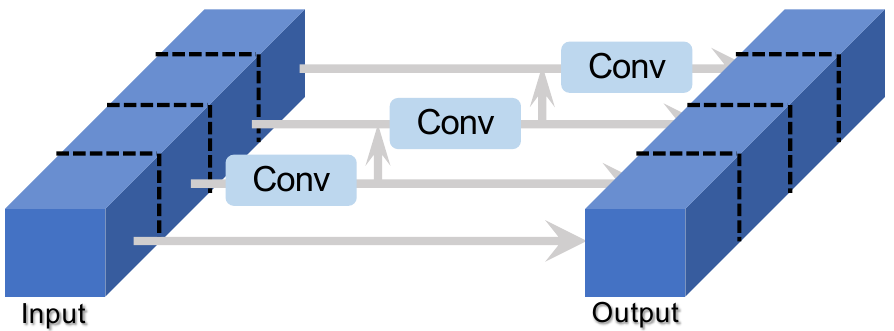}
    \caption{Res2-Conv Module Structure: The input’s feature maps are split into multiple sub-feature maps evenly along the spectral channel dimension, and the sub-feature after the convolution is added to its neighboring sub-feature.}
    \label{fig:res2-conv}
\end{figure}

\subsection{Deep Learning Based Reconstruction \label{Sec:DL_al}}

\subsubsection{Network Structure}
The optimization-based reconstruction methods typically need tens to hundreds of iterations to converge, and often require fine-tuning of the hyper-parameter (e.g., $\tau$ in ~\eqref{Eq:constrain}) to get good results, both of which are time-consuming. In addition, memory requirement is also huge due to the complex computation such as matrix inversion.
To address these issues, we developed an end-to-end deep learning framework to accelerate the reconstruction towards real time imaging and display. We adopted the U-net structure~\cite{Unet_RFB15a} (Fig.~\ref{fig:Unet}) as the backbone of our network, which features a U-shaped encoder-decoder structure (left and right side of Fig.~\ref{fig:Unet}). We set $5$ convolution stages in each side of the U-net with each stage consisting of two convolution layers (each encoder stage ends with a pooling layer and each decoder stage starts with a deconvolution layer). Five skip connections are set between stages of the same depth (gray horizontal lines in Fig.~\ref{fig:Unet}) to concatenate the output of the encoder stage and the output feature of the deconvolution layer of the decoder stage.
The activation function in the last convolution layer is set as $tanh(\cdot)$ to ensure a desired scale of the final output.

To reduce the number of required variables and to learn the relationship between different features, we introduce the Res2-Conv module~\cite{gao2019res2net} to our network. The novel CNN-based structure exploits multi-scale feature extraction ability along different spectral channels and shows better performance compared with ResNet~\cite{he2016deep}, on the localization of object activation mapping~\cite{Selvaraju_2017_ICCV}. 
As shown in Fig.~\ref{fig:res2-conv}, the input feature maps are split into multiple sub-feature maps evenly along the spectral dimension. The $i^{th}$ sub-feature incorporates the output of the $(i-1)^{th}$ convolution layer through element-wise addition and is then sent to the next convolution layer.
Then, starting from the third block in both encoder and decoder, we replace the second convolution layer in each block with the \emph{Res2-Conv} module. Notably, when the numbers of input and output channel are the same, the number of trainable parameters in Res2Net is smaller than that in a traditional convolution layer.

\subsubsection{Training}

Ideally, the input and output of the network should be set as the 2D compressed measurement and the 3D image (in the form of 2D image stacks), respectively. In practice, to reduce the learning burden, we initialize the network input as {$\Phimat \ts \yv$ with $\Phimat\ts$ being an approximate inverse operator}. Such setting is considered to be reliable for computational imaging problems~\cite{Barbastathis19DL} and has been widely used in recent works~\cite{goy2018low,lyu2017deep}. 
Besides, the network output is chosen to be the spatial-spectral datacube instead of the volumetric cube to avoid learning the inverse Fourier transform operation.

Training samples are generated by {\em simulating} the system forward encoding using the experimentally calibrated mask pattern and parameters of the hardware. 
To reduce the discrepancy between the simulated and experimental data, several real system noises are considered in the simulation as follow. 
\begin{itemize}
    \item [1)] {\em Shot noise} is introduced by simulating Poisson sampling on each pixel of the simulated measurement.
    \item [2)] {\em Discretization noise} due to CCD's discrete sampling array is introduced by first implementing a much finer spatial grid than the CCD array to generate a fine measurement, and then downsampling the measurement to the precision of the CCD array.
    \item [3)] The {\em non-uniform illumination} in the real system is considered by superposing the calibrated illumination pattern on both the reference mirror and object in the simulation.
\end{itemize}

As 3D objects are complex, we currently only use single-layer objects (with different $z$ locations) to train the network. The objects are random numbers, letters and their combinations, with various transverse offsets, rotations and axial locations. 
$20000$ sample are sent for training in which $2000$ are used for validation.
We train the model by feeding $\Phimat \ts \yv$ into the network and using the Adam optimizer~\cite{kingma2014adam} to minimize the objective function - root mean square error (RMSE) - between the model prediction and the truth $\xv$. 
For each epoch, we use measurements of the same objects but with different shot noise realizations to improve the robustness of the network. 
The leaning rate is initialized as $0.01$ and scaled down by a factor of $0.5$ after each $10$ epochs.  
The training takes $4$ weeks for data size of $1300\times 1000\times 200$ ($(x,y,\lambda)$); then experimental measurements are used for testing which take only $100ms$ each, significantly increasing the reconstruction speed of our approach.

\section{System}
\label{Sec. system}
\begin{figure*}[htbp!]
	\centering
	\fbox{\includegraphics[width=1\linewidth]{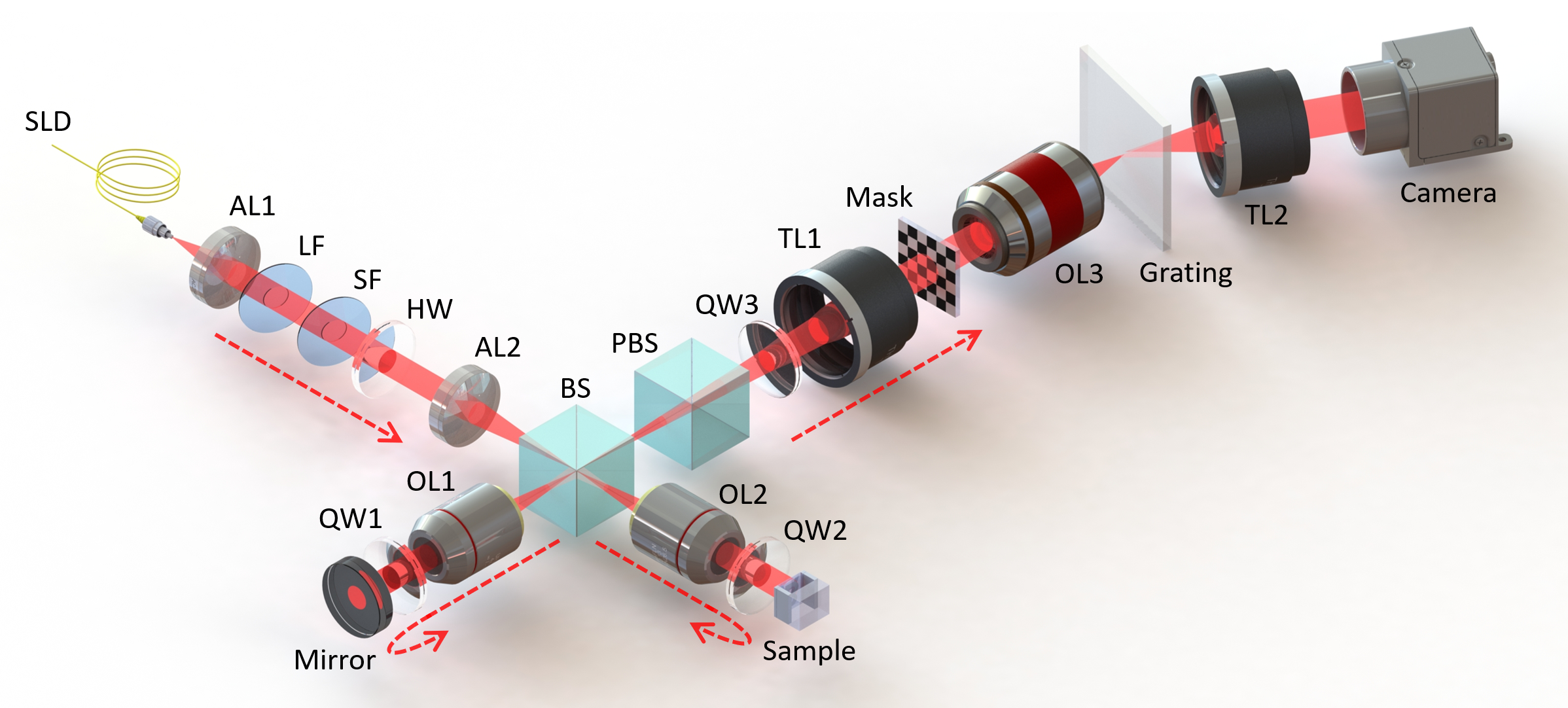}}
	\caption{Optical setup. SLD: superluminescent diode, AL: achromatic doublet lens, LF: long-pass filter, SF: short-pass filter, HW: half-waveplate, QW: quarter-waveplate, BS: beamsplitter, PBS: polarizing beamsplitter, OL: objective lens, TL: tube lens.}
	\label{fig:setup}
\end{figure*}

\subsection{Optical setup}
The optical setup of our system is shown in Fig.~\ref{fig:setup}. A superluminescent diode (SLD) source with $830nm$ center wavelength and $20nm$ bandwidth (full width at half maximum (FWHM)) is delivered to the optical setup through a single-mode fiber. The light from the fiber is collimated by an achromatic doublet lens AL1 into a beam of $5 mm$ in diameter. The beam is then splitted and relayed onto the reference mirror and the object through two identical 4-f systems composed of a common achromatic doublet lens AL2 and two objective lenses OL1 and OL2 (with varying focal length depending on the required lateral FoV). Images of the reference mirror and the object are then relayed and combined coherently through another two identical 4-f systems consisting of a common tube lens TL1 ($f=100 mm$) and two objective lenses OL1 and OL2. Thus, a depth-encoded spatial-spectral datacube is formed before the coded aperture, a mask applying binary random amplitude modulation. The modulated spatial-spectral datacube after the mask is further relayed to the camera (Edge 5.5, PCO, Germany, $2560\times 2160$ pixels resolution, $6.5\mu m$ pixel pitch) by another 4-f system composed of an objective lens OL3 (2X, 0.1NA) and a tube lens TL2 ($f=100 mm$). A grating (Wasatch Photonics, USA, $600lp/mm$) is inserted between OL3 and TL2 to introduce different lateral shifts to different spectral channels of the modulated spatial-spectral datacube before it collapses onto the camera plane.

\begin{figure*}[htbp!]
	\centering
	\fbox{\includegraphics[width=1\linewidth]{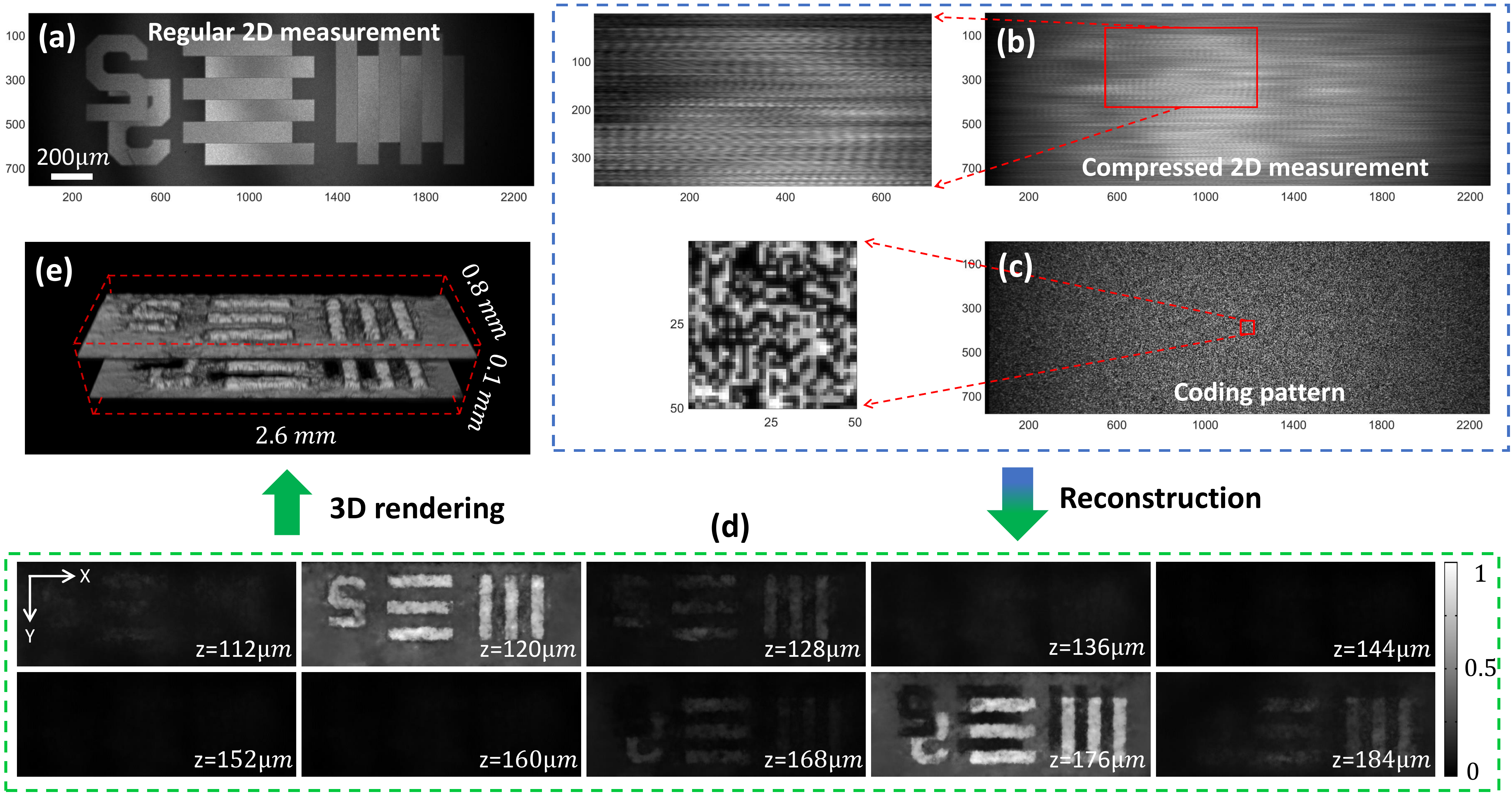}}
	\caption{Experimental results of a double layer phantom. CS algorithm is used for reconstruction. The object is created by stacking two resolution target plates with their patterns laterally staggered. (a) A 2D image of the sample captured by a regular 2D camera.  (b) Compressed measurement of the object captured by our system. (c) Image of the coded aperture. (d, e) 3D image of the object reconstructed from (b) and (c) shown in the form of volumetric rendering and cross sections, respectively.}
	\label{fig:2layer}
\end{figure*}

\begin{figure*}[htbp!]
	\centering
	\includegraphics[width=1\linewidth]{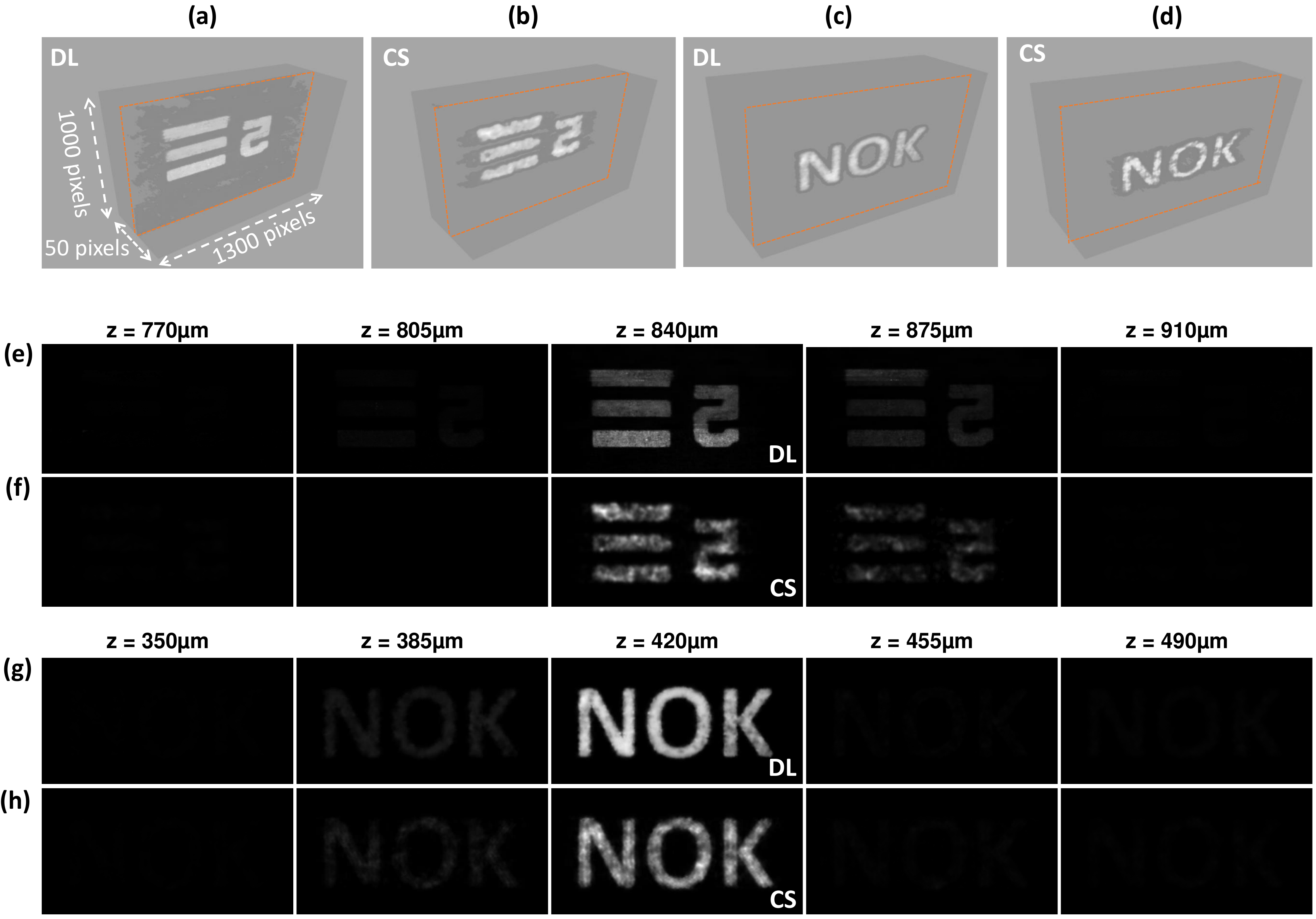}
	\caption{\label{fig:dl_results} Comparison of reconstruction using CS and DL algorithms. Two objects with different patterns and z locations are tested. (a, b, e, f) are results of a standard resolution target (a subregion with a number 5 and three horizontal bars is chosen). (c, d, g, h) are results of a `NOK' pattern (chrome pattern plated on quartz substrate). The shown volumes have a size of $1000\times 1300\times 50$ ($5$ $x$-$y$ cross sections shown in (e-h) in linear gray scale). See Multimedia 1 for more comparison between CS and DL reconstruction.}
\end{figure*}

\begin{figure*}[htbp!]
	\centering
	\fbox{\includegraphics[width=1\linewidth]{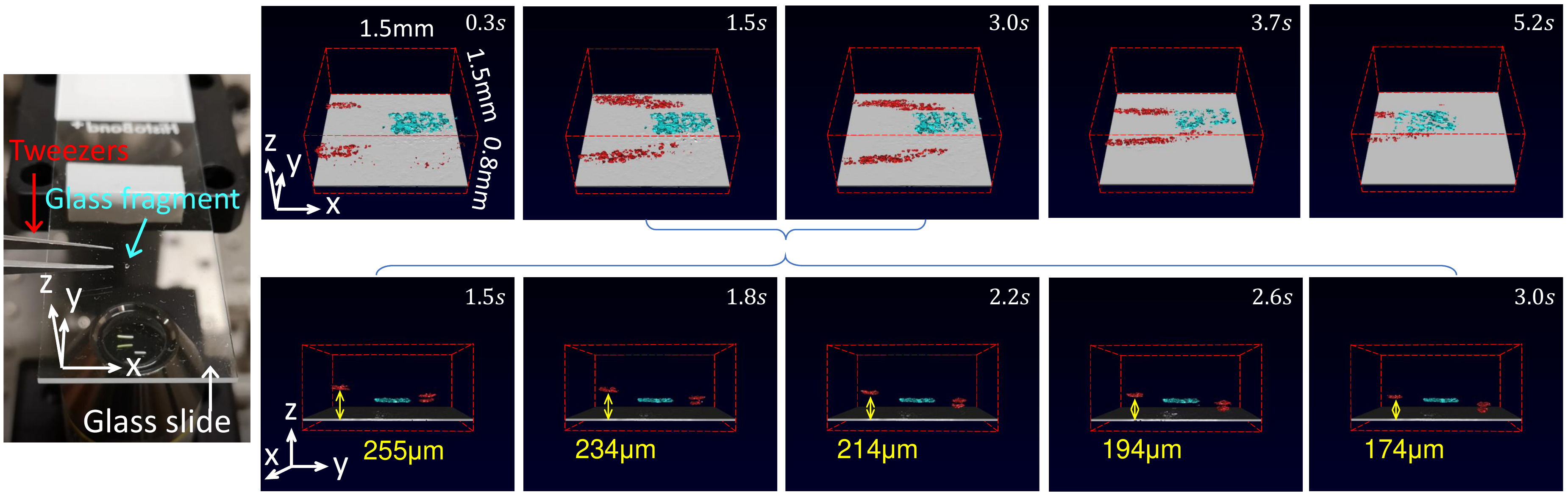}}
	\caption{4D (temporal volumetric) visualization of tweezers motion - picking up a small glass fragment from a glass slide (recorded for $6$ seconds at $50 fps$, see  Multimedia 2 for the complete video). A photo of the scene is shown on the left-hand side. The motion consists of four steps: moving the open tweezers towards the glass fragment along the x-axis, descending the tweezers along the z-axis to the same depth level as the glass fragment, pinching the tweezers and moving out the glass fragment. The first row shows five representative excerpts from the complete video in the form of volumetric renderings. The second row depicts the volumes from a side view to show the depth change of the tweezers.}
	\label{fig:tweezers}
\end{figure*}

A long-pass and a short-pass filter are included before the interferometer. Both filters have a $850nm$ nominal cut-off wavelength, but by tilting their angle of incidence we can tune the cut-off wavelength and thus the bandwidth of the SLD source. Given a fixed spectral resolution, different compression ratios can therefore be achieved (see Appendix~\ref{Sec:cr}). The spectral resolution $\delta_\lambda$ of the system is determined by
\begin{equation}
    \delta_\lambda=\frac{\Delta_pd}{f},
    \label{Eq:spectral resolution}    
\end{equation} 
where $d$ is the grating constant, $\Delta_p$ is the camera pixel pitch and $f$ is the focal length of TL2. Given $\Delta_p=6.5\mu m$, $d=1.66\mu m$ and $f=100mm$, the spectral resolution is calculated to be $0.1nm$. According to Eq.\eqref{Eq:depth fov}, this corresponds to a maximum imaging depth of $1.6mm$ in air.

We find that the stray light in the system, primarily from the surface reflection of BS, OL1 and OL2, would contaminate or even dominate the signal when imaging objects with low reflectivity. To address this issue, we employ a polarization-selective isolator consisting of several waveplates and a polarizing beamsplitter (PBS). The isolator labels the signal, \ie, reflection from the object and the reference mirror, and the stray light, \ie, reflection from BS, OL1 and OL2, with P- and S-polarized states, respectively, and then deflects the stray light off the optical path with the PBS (see Appendix~\ref{Sec:isolator} for detailed description).

Note that the relatively narrow spectral width of the SLD to white light as used in previous hyperspectrometers~\cite{wagadarikar2008single} allows us to use a grating rather than a prism as the dispersive element without suffering from space crosstalk between diffraction orders of the grating. One benefit of the grating is its better linearity between dispersion angle and wavelength, which allows us construct an accurate 3D sensing matrix from a single image of the coded aperture without additional calibration effort~\cite{wagadarikar2009video}.

\subsection{System validation}
In this section, we validate our system using both the developed CS and DL reconstruction algorithms.

We fabricated a double-layer phantom by stacking two 1951 USAF resolution targets with their patterned surfaces separated by $50\mu m$ in z-axis and $110\mu m$ in both the two lateral dimensions, as shown in Fig.~\ref{fig:2layer}(a). The raw 2D compressed measurement of the sample is shown in Fig.~\ref{fig:2layer}(b), which features interference fringes and smearing effect due to the dispersion along the horizontal dimension. Fig.~\ref{fig:2layer}(c) shows the calibrated image of the coded aperture (see Appendix~\ref{Sec:cali} for calibration method). As mentioned in Sec. \ref{Sec. method}, the DC terms in the measurement should be subtracted before performing the reconstruction. The DC terms $I_R$ and $I_S$ are captured by blocking the reference or object arm, respectively, with all other conditions the same as when capturing the measurement. 3D reconstruction of the object using the developed CS algorithm is shown in Fig.~\ref{fig:2layer}(d) in the form of lateral cross-sections ($10$ out of $200$ shown). The interval between adjacent cross-sections is $8\mu m$, which is determined by the selected spectral width in the reconstruction, rather than the FWHM of the SLD as used to calculated the axial resolution in Eq.\ref{Eq:depth resolution}, see Appendix~\ref{Sec:Depth Interval}). We can see from the results that the axial positions of the two patterned surfaces are sharply resolved at $z=120\mu m$ and $176\mu m$, respectively, and the lateral structures of the object are also well resolved with sharp edges compared to the smeared measurement shown in Fig.~\ref{fig:2layer}(b). We also observe a negative pattern at $z=176\mu m$, which is the shadow cast by the top layer. Fig.~\ref{fig:2layer}(e) shows a volumetric render of the reconstructed 3D object.

In this experiment, the two filters are tilted to achieve a $40nm$ bandwidth (resolved spectral width, rather than FWHM spectral width, see Appendix~\ref{Sec:Depth Interval}). Given the $0.1nm$ spectral resolution from Eq.\eqref{Eq:spectral resolution}, this bandwidth corresponds to a $400$ compression ratio in the measurement. In other words, we have resolved  $400$ spectral channels through CS reconstruction with a snapshot measurement. However, after performing the inverse Fourier transform along the spectral dimension, half of the cross sections, which are mirror images as mentioned in Sec.\ref{Sec. method}, are discarded, leaving $200$ cross sections with $8\mu m$ interval in the final 3D image. Also note that due to the unshearing operation after the CS reconstruction and before the inverse Fourier transform (See Fig.~\ref{fig:principle}), the $2560\times2160$ pixels measurement gives a $2160\times2160$ lateral size of the reconstructed 3D image, with $400$ pixels (the compression ratio) lost in the horizontal/dispersion dimension. Therefore, the reconstructed voxel number from one measurement is $2160\times2160\times200\approx1G$. The results shown in Fig.~\ref{fig:2layer} are obtained after $50$ iterations of the CS reconstruction algorithm, which takes $30$ minutes on a computer with $12$ CPU cores @ $3.2$ GHz and $64$G memory. 
In the following, we show reconstruction using the fast deep learning network developed in Sec.~\ref{Sec:DL_al} which, as mentioned before, only takes $100ms$ per measurement.

A thorough characterization of our system performance including spatial and axial resolution and sensitivity can be found in Appendix~\ref{Sec:performance}.

\subsection{Deep learning reconstruction results}
As mentioned in Sec.~\ref{Sec:DL_al}, due to the complexity of 3D objects, we only used simulated single-layer objects (with different depths) to train and validate our network. We now test the network using {\em real measurements} of single-layer objects, which did not appear in the training but has similar structural features to the data used in the training (which are numbers and letters, see Sec.~\ref{Sec:DL_al}). We choose two objects with different transverse patterns and axial locations for the testing. The first object is a standard resolution target (a subregion with a number `5' and three horizontal bars). The reconstruction from the CS and DL algorithms is shown in Fig.~\ref{fig:dl_results} in the form of volumetric renderings (a,b) as well as $x$-$y$ cross sections (e,f). The second object is a `NOK' pattern (chrome pattern plated on quartz substrate) and the results are shown in Fig.~\ref{fig:dl_results} (c,d) and (g,h).
For each measurement, the testing takes $100ms$ on GPU and the subsequent inverse Fourier transform takes $50ms$ on CPU.
From the results we see that for this specific type of object, the deep learning algorithm is able to provide better image quality than the CS algorithm both in terms of the intensity uniformity and structural sharpness. 
The expansion of the image along the $z$-axis (e.g., see the dim patterns at $z=805\mu m$ and $875\mu m$ in Fig.~\ref{fig:dl_results} (e,f)) is because that the real objects are not placed exactly at the discrete resolved $z$ planes determined by the fast inverse Fourier transform. 
For more comparison between the CS and DL algorithms, please see Multimedia 1. 
These preliminary results of single-layer objects demonstrate the feasibility of adopting the DL method to our imaging system, which is challenging due to the large data size ($1300\times 1000\times 200$ in $(x,y,\lambda)$ for the presented results) and the complex system forward operator that encompasses the encoding of the interferometer and the compression process. 
Based on these promising results, our next work is to incorporate more complex 3D objects to train the network.

\begin{figure*}[htbp!]
	\centering
	\fbox{\includegraphics[width=1\linewidth]{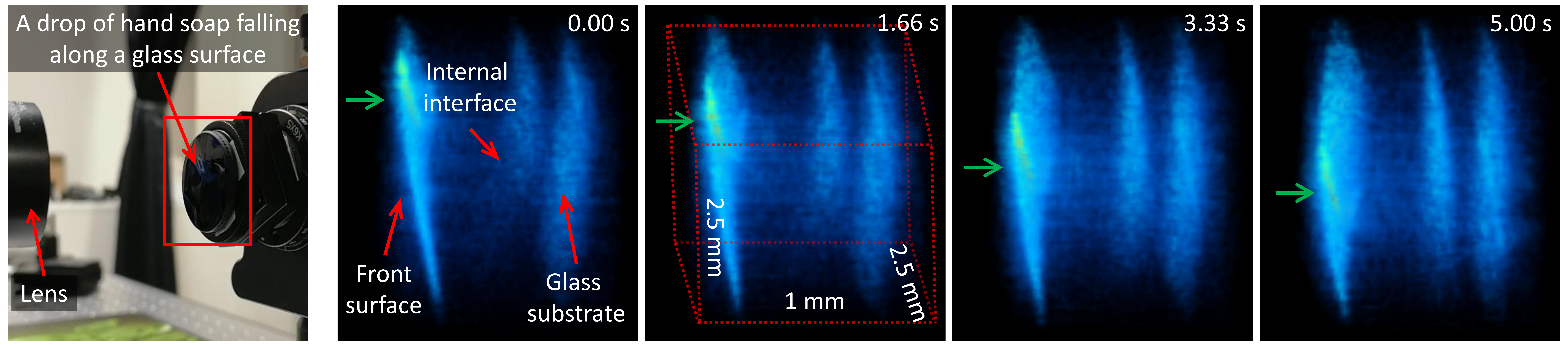}}
	\caption{4D visualization of a falling liquid drop (recorded for $5$ seconds at $100fps$, see Multimedia 3 for the complete video). The liquid is a drop of hand soap falling along a glass surface. A photo of the scene is show on the left-hand side. Four representative excerpts are shown in the form of volumetric renderings. The green arrows show the change of the front surface (convex curve) of the droplet. The internal interface is caused by the high viscosity of the hand soap. Pseudo-color is used for better visualization.}
	\label{fig:liquid}
\end{figure*}
\section{Capture dynamic 3D scenes }
\label{Sec:3D_result}

The single-shot feature of our approach allows us to capture dynamic 3D scenes at video rates, \ie, frame rates of regular 2D cameras. We now show this through several examples. In the first experiment, we used tweezers to pick up a small glass fragment, which simulates industrial inspection or repairing process. A photo of the scene is shown on the left-hand side of Fig.~\ref{fig:tweezers}. The sample arm in this experiment was setup vertically to simplify the operation. The glass fragment had a size of $0.4\times0.8\times0.2mm$ and was placed on top of a glass slide. The whole procedure included four phases: i) move the open tweezers towards the glass fragment along the $x$-axis, ii) descend the tweezers along the $z$-axis to the same depth level as the glass fragment, iii) pinch the tweezers and iv) move out the glass fragment. The scene was recorded at $50fps$ for $6s$. Five representative excerpts from the total $300$ reconstructions (see Multimedia 2 for the complete video) are presented in Fig.~\ref{fig:tweezers} (first row) in the form of volumetric renderings. The second row in Fig.~\ref{fig:tweezers} depicts the rendered volumes from a side view to show the depth change of the tweezers. The speckle artifacts in the images are due to the rough surfaces of the tweezers and the glass fragment which cause non-uniform back-scattering.

\begin{figure*}[!]
	\centering
	\fbox{\includegraphics[width=1\linewidth]{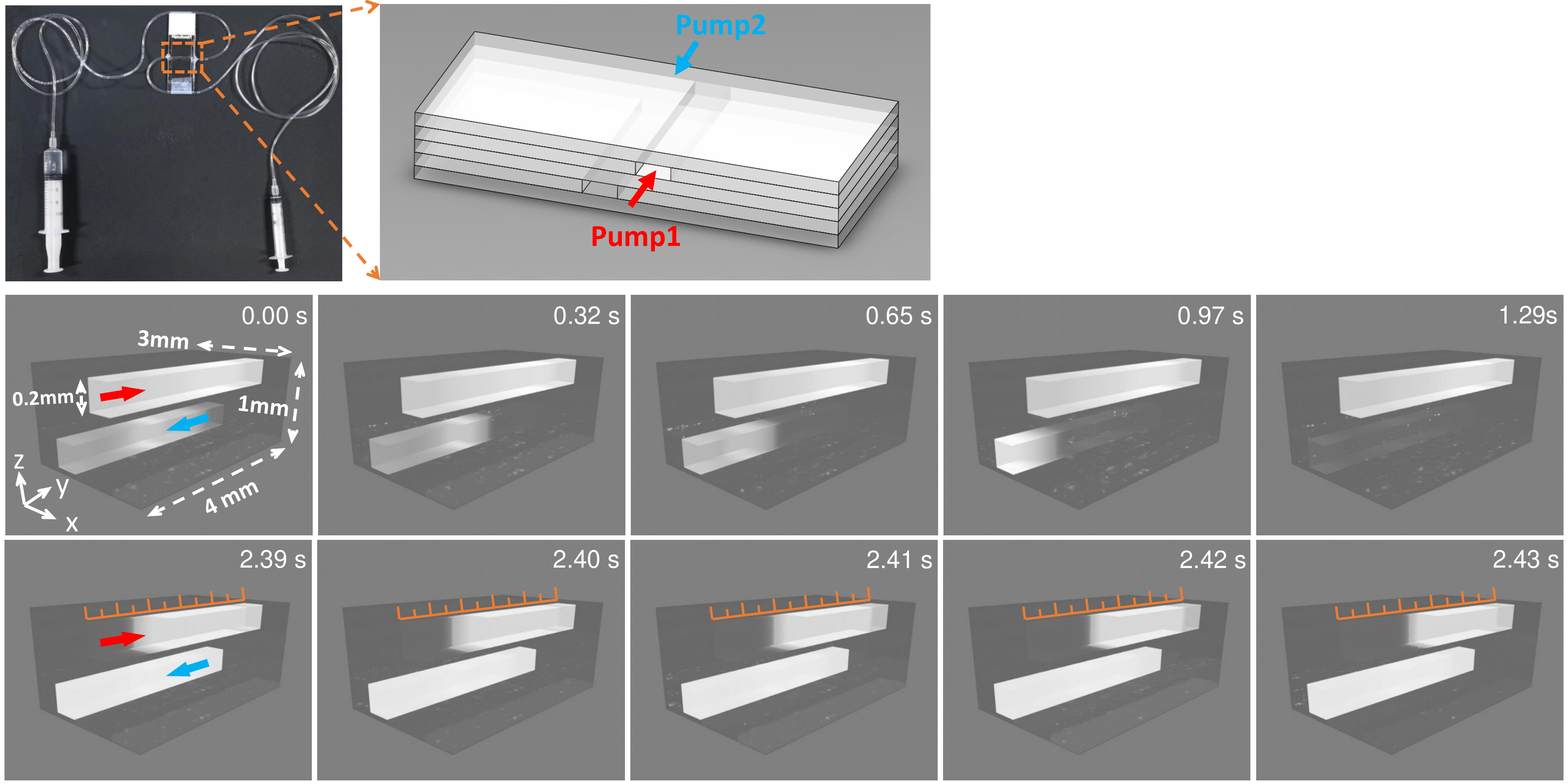}}
	\caption{4D visualization of flowing water in closed tubes (recorded for $3$ seconds at $100fps$, see Multimedia $4$ for the complete video). A photo of the sample is shown in the first row (left). The tubes are created from multiple pieces of glass slides of $200 \mu m$ thickness (see the diagram in the first row (right)). Water with air bubbles is pumped into the two channels from different directions using syringes. $10$ representative excerpts from total $300$ reconstructions are shown in the second and third rows in the form of volumetric renderings.}
	\label{fig:flow}
\end{figure*}

\begin{figure*}[!]
	\centering
	\fbox{\includegraphics[width=1\linewidth]{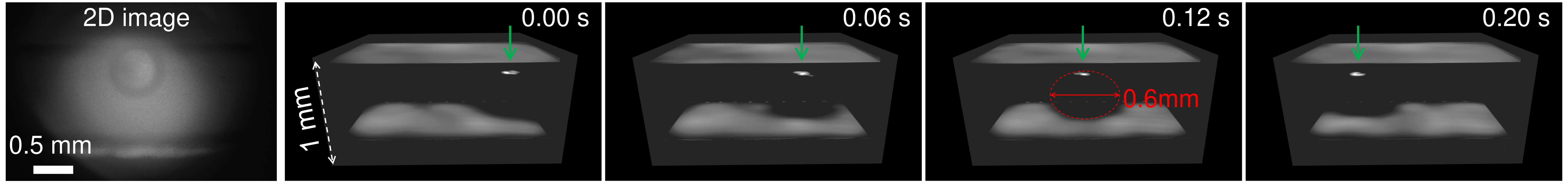}}
	\caption{4D visualization of a particle flowing in a water tube (recorded for $0.2$ seconds at $100fps$, see Multimedia 5 for the complete video). The particle, glass ball with aluminum coating, $0.6mm$ in diameter, are pumped into a glass tube of $1\times 20\times 1mm$ in $(x,y,z)$ with a syringe. A 2D image of the tube when the particle is in the middle of the tube is shown on the left. Four excerpts from the total 20 reconstructions are shown in the form of volumetric renderings.}
	\label{fig:particle}
\end{figure*}

The second scene we show is a drop of liquid (hand soap) falling along a glass surface. A photo of the scene is show on the left-hand side of Fig.~\ref{fig:liquid}. This time the scene was recorded by our system at $100fps$ for $5s$. Four representative excerpts from the total $500$ reconstructions (see Multimedia 3 for the complete video) are shown in Fig.~\ref{fig:liquid} in the form of volumetric rendering. The green arrows show the change of the front surface (convex curve) of the droplet. We observe that an internal interface was formed (and changing) during the falling of the droplet. This is due to the high viscosity of the hand soap. This example also demonstrates that our approach is able to resolve multi-layer semi-transparent objects. The bright background in the reconstructed volume arises from inside back-scattering due to the inhomogeneity of the liquid.

The third scene we show here is flowing water in closed tunnels. The tubes were created by stitching multiple pieces of glass slides (see the diagram in the first row of Fig.~\ref{fig:particle}), which have a size of $1\times 20\times0.2$ mm with different depth. Water with air bubbles was pumped into the two tubes from opposite directions using syringes. To capture a larger transverse FOV, the objective lenses OL1 and OL2 in Fig.~\ref{fig:setup} were replaced by two achromatic doublet lenses with $50$ mm focal length, giving a FoV of $7\times 7\times 1.6$ mm. The motion of the fluid was recorded for $3$ seconds at $100$ fps with $1400\times 1000$ pixels transverse image size (effective sensor size). Ten representative frames from total $300$ reconstructions (Multimedia 4) are shown in the second row of Fig.~\ref{fig:particle} in the form of volumetric renderings. Note that since water and air are both optically homogeneous, there is no reflection (and thus no signal) from the inside of the tubes. In order to better visualize the motion of the fluid front, the voxels inside the tubes have been rendered with the average value of the signal from the tube surfaces. Also note that the darker parts of the tubes represent the water and the lighter parts the air, because water has a refractive index closer to glass than air and therefore causes less light reflection from the interfaces.

The last scene we show is a particle flowing in a water tube. The tube was created using the same method as in the third scene. Water and particles ($0.6mm$ in diameter) was pumped into the tube through a syringe.  A photo of the tube with a particle in the middle is shown on the left-hand side of Fig.~\ref{fig:particle}. The motion was recorded for $0.2s$ at $100fps$ when a particle entered the FoV. Four representative frames from the complete video (see Multimedia 5) are shown in Fig.~\ref{fig:particle}. Due to the high curvature of the particle, only the reflection from the top of the particle (indicated by the red circle in Fig.~\ref{fig:particle}) was collected by the objective lens. Therefore, the particle appears as a bright spot floating above a round shadow of the same diameter as the particle in the 3D renderings.

\section{Discussion}
\label{Sec:con}
We have demonstrated a single-shot compressive 3D imaging system which employs an interferometer for depth sensing and a snapshot spectrometer for fast sampling. Both a compressive sensing and a deep learning algorithm are developed for image reconstruction.
The interference-based depth sensing scheme enables us to achieve a uniform depth resolution of $13\mu m$ across an axial FoV of $1.6mm$. The image-space coding regime, as opposed to the pupil coding regime, dramatically reduces the computation complexity and enables us to reconstruct $1G$ voxels from a 2D compressed measurement with a med-end computer.
Due to the fast imaging speed and micrometer-scale depth resolution, we expect our approach to find applications in optical metrology, machine vision, fluid dynamics, \etc

A trade-off exists between the lateral and axial throughput in our approach due to the compression of the axial information into the lateral dimension. As shown in Appendix~\ref{Sec:performance} (Fig.~\ref{fig:lateral_resolution}), higher compression ratios, which mean higher axial throughput, lead to lower lateral resolution and thus lower lateral throughput given a fixed lateral FoV. This trade-off can be mitigated by developing more advanced reconstruction algorithms and/or employing more relevant priors according to the structural features of the dataset.

The $40$dB sensitivity of our approach (see Appendix~\ref{Sec:performance} Fig.~\ref{fig:axial_resolution}) prevents our approach from imaging deep into biological samples due to the weak back-scattering. However, as demonstrated in Sec.~\ref{Sec:3D_result}, this sensitivity is nevertheless far sufficient to detect industrial samples such as metals, glasses and semiconductor materials. Besides, like the spatial resolution, we believe the sensitivity can also be improved by employing more advanced/dedicated algorithms.

Our follow-up work would be incorporating more complex 3D objects in training the deep learning network. A digital micromirror device might be used to generate the training data instead of using the simulation data, which reduces the discrepancy between the training and testing data. 3D objects can be simulated by synthesizing multiple measurements of 2D objects at different axial locations.

\appendices

\section{Characterization of System Performance}
\label{Sec:performance}

In this section, we characterize the performance of our system including lateral and axial resolution and sensitivity. As all the characterization is performed with a single-layer object (either a resolution target or a mirror), both the earlier developed CS and DL algorithms can be used for reconstruction. However, since our current DL network has not been generalized to arbitrary 3D objects, it might be unfair to use DL for the characterization though it provides better results than CS for single-layer objects. For this reason, all the following metrics are evaluated based on CS reconstruction.

\subsection{Lateral resolution}

\begin{figure*}[t]
	\centering
	\fbox{\includegraphics[width=1\linewidth]{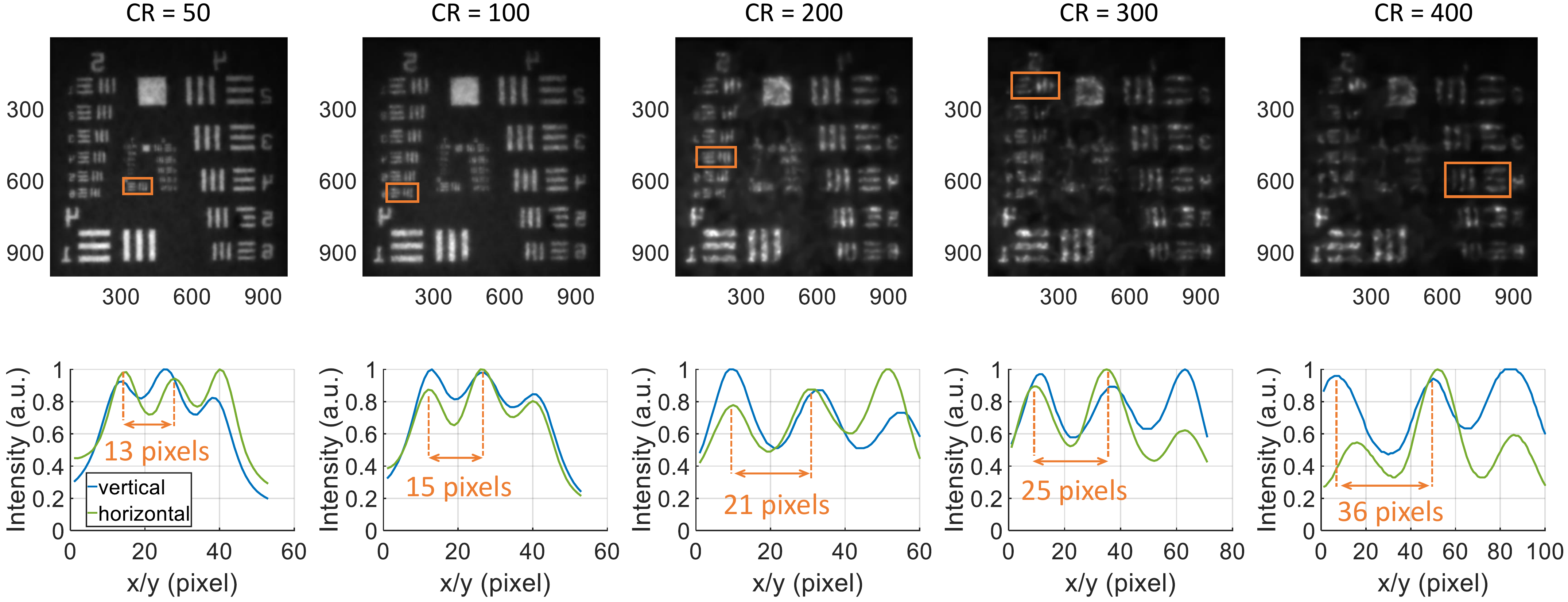}}
	\caption{Experimental evaluation of lateral resolution under various compression ratios (CR).}
	\label{fig:lateral_resolution}
\end{figure*}

In our system, the mask feature size, \ie, size of individual code element of the mask, is bigger than the diffraction-limited resolution of both the 4-f relay systems before and after the mask. Therefore, the mask feature size poses the limit to the lateral resolution. To be more specific, the minimal resolvable spatial period is two mask elements, scaled by the magnification of the pre-relay 4-f systems (the one between the mask and the object, see Fig.~\ref{fig:setup}). However, in practice, this 2-element limit is hard to reach in any CS-based imaging systems due to the imperfect reconstruction, which is caused by many factors such as measurement noise (mainly shot noise), background noise (ambient light), calibration errors and the limited reconstruction capability of the CS algorithms. 
As such, the real lateral resolution in CS-based imaging systems is typically several or tens of mask elements.
We use a standard resolution target (1951 USAF) as the object to evaluate the lateral resolution of our system. The line pairs in the reconstructed image are considered resolvable if the reconstruction has a dip of at least 20\%, as in the Rayleigh criterion.
The results for various compression ratios (CR) are shown in Fig.~\ref{fig:lateral_resolution}.
We can see that the resolution increases (becomes worse) with CR, which is because higher CR poses more challenge to the reconstruction algorithm.
The orange rectangles in the upper part of Fig.~\ref{fig:lateral_resolution} indicate the finest distinguishable line pairs, and their averaged 1-D vertical and horizontal profiles are depicted in the lower part.
The lateral resolution is determined to be $6.5$, $7.5$, $10.5$, $12.5$ and $18$ mask elements for CR = $50$, $100$, $200$, $300$ and $400$, respectively. 
As mentioned above, even for the smallest CR of $50$, the lateral resolution is still $3$ times larger than the 2-element limit.
However, encouraged by the better results from deep learning reconstruction shown in Fig.~\ref{fig:dl_results}, we anticipate a well trained neural network to provide a lateral resolution close to the theoretical limit.

\begin{figure*}[htbp!]
	\centering
	\fbox{\includegraphics[width=1\linewidth]{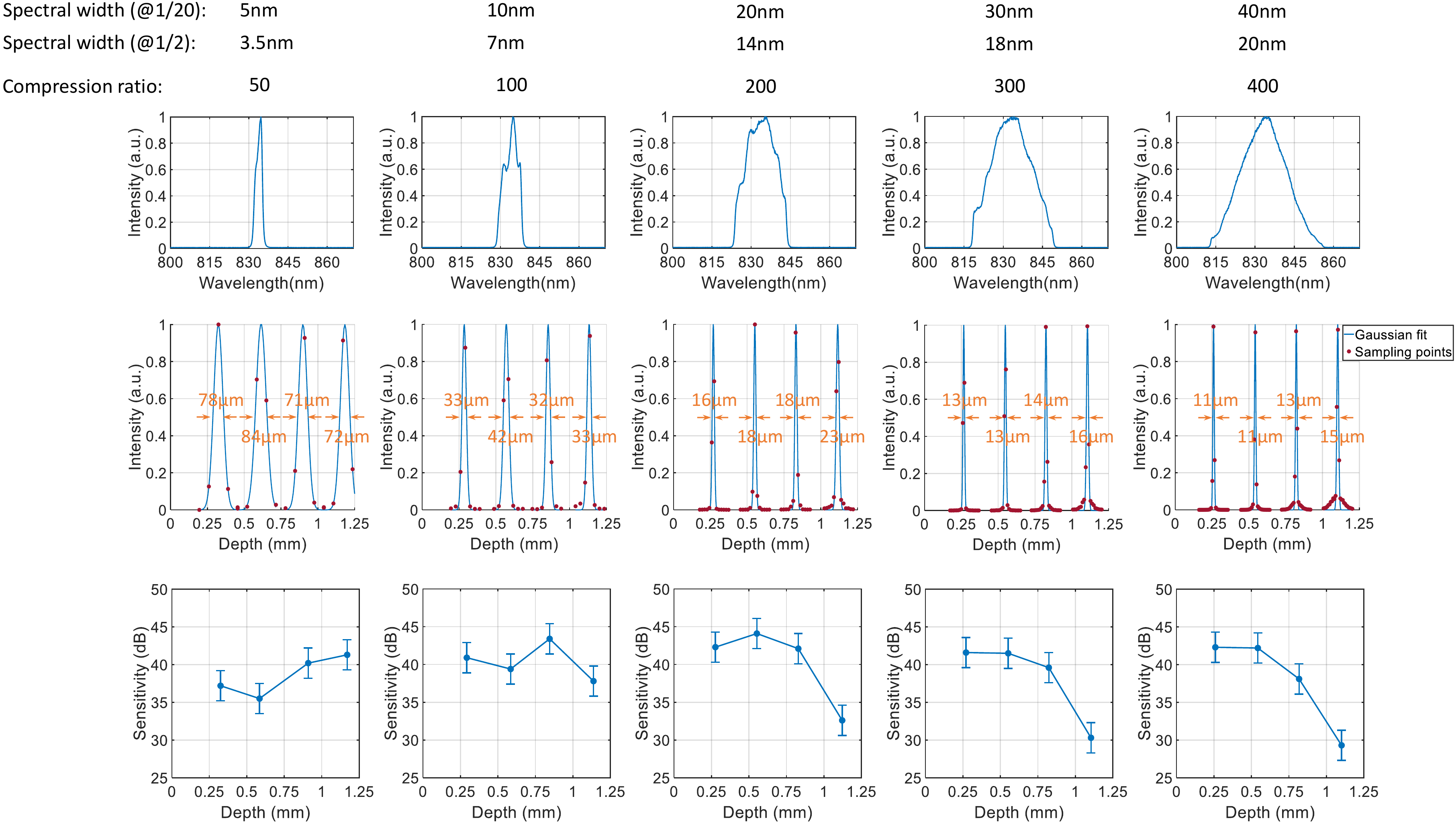}}
	\caption{Experimental evaluation of axial resolution (second row, Appendix.~\ref{Sec:axial_Res}) and sensitivity (third row, Appendix.~\ref{Sec:sen}) under various compression ratios.}
	\label{fig:axial_resolution}
\end{figure*}

\subsection{Axial resolution \label{Sec:axial_Res}}

The axial resolution of our system is evaluated by using an object with infinitesimal axial width, a mirror, to measure the axial point spread function~\cite{dubois2004ultrahigh} of our system. The 1D axial profile of the mirror is acquired by averaging the reconstructed 3D image transversely. Then a Gaussian curve is fit to the profile and the FWHM of the Gaussian curve is taken as the axial resolution. For compression ratios of $400$, $300$, $200$, $100$ and $50$, which correspond to FWHM spectral widths of $20$, $18$, $14$, $7$ and $3.5nm$, the axial resolution (second row in Fig.~\ref{fig:axial_resolution}) is measured to be $13$, $15$, $19$, $37$ and $78\mu m$, respectively, which are consistent with the theoretical prediction of $15$, $17$, $22$, $43$ and $87\mu m$ from ~\eqref{Eq:depth resolution}.

\subsection{Sensitivity \label{Sec:sen}}
The sensitivity of an imaging system describes the capability to detect weak signal. In our system, it is defined as the sample arm reflectivity at which the signal-to-noise ratio (SNR) drops to unitary~\cite{leitgeb2003performance}. Numerically, the sensitivity is equal to the SNR measured when using a mirror of $100\%$ reflectivity as the object.
For a regular Fourier domain optical coherent tomography system, the theoretical sensitivity $\Sigma_{FD}$ is expressed by~\cite{leitgeb2003performance}:
\begin{equation}
\Sigma_{FD}=\frac{\frac{1}{N}\left(\frac{\rho \eta \tau}{h\nu_0}P_0\right)^2\gamma_s\gamma_rR_r}{\frac{\rho \eta \tau P_0}{h\nu_0N}\gamma_rR_r\left[1+\frac{(1+\Pi^2)\rho \eta P_0}{2h\nu_0N}\gamma_rR_r\frac{N}{\Delta \nu_{eff}}\right]+\sigma^2_{receiver}},
\label{Eq. sensitivity1}
\end{equation}
where $P_0$ is the power of the source, $\rho$ is the diffraction efficiency of the grating,  $\eta$ and $\tau$ are the quantum efficiency and exposure time of the detector, respectively; $N$ is the number of discrete wavelengths or the number of pixels of the 1D detector array, $h$ is the Planck constant, $\nu_0$ is the central frequency of the source, $\gamma_r$ and $\gamma_s$ are parts of the input power that exits the interferometer through the reference and the sample arm, respectively;  $R_r$ is the reflectivity of the reference mirror, $\Pi$ is the degree of polarization, $\nu_{eff}$ is the effective bandwidth of the source, and $\sigma^2_{receiver}$ is the receiver noise of the detector which consists of dark current noise and readout noise, \ie, $\sigma^2_{receiver}=\sigma^2_{dark}+\sigma^2_{readout}$. In shot noise limit, $\sigma^2_{receiver}$ can be ignored (for our camera, $\sigma^2_{dark}=0.6 e^-/pixel/s$, $\sigma^2_{readout}=1.0e^- med /1.4e^- rms$). Then ~\eqref{Eq. sensitivity1} is simplified to:
 \begin{equation}
\Sigma_{FD}=\frac{\frac{\rho \eta \tau}{h\nu_0}P_0 \gamma_s}{1+\frac{\rho \eta}{h\nu_0}P_0\gamma_rR_r\frac{1}{\Delta \nu_{eff}}}.                           
\end{equation}
Given $\rho=80\%$, $\eta=35\%$, $P_0=10mW$, $R_r=1$, $\gamma_r=0.5$, $\nu_o=3.6\times10^{14}/s$ and $\nu_{eff}=8.7\times10^{12}/s$, we have $\frac{\rho \eta}{h\nu_0}P_0\gamma_rR_r\frac{1}{\Delta \nu_{eff}} \ll 1$. Then we obtain
\begin{equation}
\Sigma_{FD}=\frac{\rho \eta \tau}{h\nu_0}P_0 \gamma_s.                           
\end{equation}
This is exactly the number of photoelectrons generated in all the $N$ detector pixels when a mirror of $100\%$ reflectivity is placed at the sample arm and the reference arm is blocked. Therefore, $\Sigma_{FD}$ is limited by the full well capacity (FWC) of the detector. However, due to the compression in our system, all the spectral signal from a given $(x,y)$ spot on the object is compressed into a single pixel of the camera, as opposed to spread into $N$ pixels. Therefore, the FWC of a single pixel, rather than $N$ pixels, should be used to calculate the sensitivity. Given $FWC=30000e^-$ for our camera, the theoretical sensitivity of our system (dubbed SI3D) in dB is calculated to be
\begin{equation}
\Sigma_{FDCS}=10\log(FWC)\approx45dB,
\label{Eq.sensitivity}
\end{equation}
which means a minimal detectable sample arm intensity reflectivity of $10^{-4.5}$.

The sensitivity of our system is evaluated by using a mirror of approximate $100\%$ reflectivity as the object. The sensitivity is calculated as twenty times the base-10 logarithm of the ratio between the axial peak value to the standard deviation of ambient noise floor~\cite{choma2003sensitivity}. Due to imperfect experimental conditions, the measured sensitivity (third row in Fig.~\ref{fig:axial_resolution}) is slightly lower than the theoretical value. The results also verify that the sensitivity of our system is independent of the compression ratio. 

\section{System Calibration}
\label{Sec:cali}
Two parameters need to be calibrated for precise reconstruction. 
The first one is the pattern of the coded aperture. We design the aperture pattern as pseudo-randomly distributed binary squares, with each square (either open or block) taking 2x2 pixels on the camera. Instead of directly using this pre-designed pattern (generated by Matlab) as the coding pattern in the reconstruction, we experimentally measure the image of the aperture on the camera. This is performed by 
\begin{itemize}
    \item blocking the sample arm, 
    \item switching the source from the broadband source to a single-frequency laser at $830nm$ (the center wavelength of the broadband source),
    \item inserting a rotating diffuser (between HW and AL2) to destroy the spatial coherence of the laser beam and obtain a uniform illumination on the aperture, and
    \item capturing the aperture pattern using the same camera as in capturing the compressed measurement.
\end{itemize}
This calibrated pattern leads to more accurate reconstruction than directly using the design pattern because imaging aberration in the real system, which is inevitable, is taken into account. 

The second system parameter that needs to be calibrated is the relative spectral position between the laser used in calibration, and the broadband source (SLD) used in capturing the compressed measurement. Their spectrum is measured by converting the present system to a spectrometer by removing Al2, TL1, OL3 and the mask, and blocking the sample arm. The laser illumination will result in a spot on the camera due to its narrow spectrum, and SLD illumination will result in a line on the camera. For each wavelength within the spectrum of the SLD, we determine its corresponding coding pattern by shifting the calibrated pattern (corresponding the laser wavelength) by a number of pixels equal to the spectral distance between the selected wavelength and the laser. These shifted coding patterns are then sent to the CS algorithm to reconstruct their corresponding spectral channels.

\section{Principle of the Isolator}
\label{Sec:isolator}
The stray light in our system, which is primarily contributed by the surface reflection of BS, OL1 and OL2, would contaminate or even dominate the signal (measurement) when imaging low-reflection objects. To isolate the stray light, we insert
\begin{itemize}
    \item a half-wave plate before the interferometer,
    \item two quarter-wave plates immediately before the reference mirror and the sample, respectively, and
    \item a polarizing beamsplitter (PBS) after the interferometer (see Fig.~\ref{fig:setup} in the main text).
\end{itemize}
The half-wave plate, whose fast axial is orientated $45$ degrees to the horizontal direction, is used to rotate the horizontally polarized light from the fiber to vertical polarization status before it enters the interferometer. The stray light reflected from the surfaces of the beamsplitter and the objective lenses remain in vertical polarization status and therefore will be deflected off the optical path when impending into the PBS. By contrast, the signal light reflected from the reference mirror and the sample is rotated back to horizontal status because it passes through the quarter-wave plates (also $45$ degrees orientated) twice, which is equivalent to passing through a $45$ degrees orientated half-wave plate once; therefore, it will be transmitted through the PBS. Note that the light reflected from the rear surfaces (closer ones to the reference mirror or the sample) of the quarter-wave plates also passes through the quarter-wave plates twice. Therefore, we slightly tilt the quarter-wave plate to prevent this reflection from being captured by the camera.

\section{Determination of System Parameters}
\subsection{Compression Ratio}
\label{Sec:cr}

In the experiments described in the main text, full width at $1/20$ maximum of the SLD spectrum (after the filters) is selected for reconstruction (by sending their corresponding coding patterns into the algorithm). This threshold is chosen as a balance between the source noise and the compression ratio. The source noise refers to the portion of the measurement that is contributed by the rest part of the spectrum, which is not considered by the algorithm and thus treated as noise. The compression ratio in the reconstruction is determined by the ratio of the chosen spectral width to the spectral resolution (see the main text for the calculation of the spectral resolution). More specifically, the compression ratio is the number of spectral channels reconstructed from a single measurement. Note that after the inverse Fourier transform, the number of depth cross sections is half of the compression ratio.

\subsection{Depth Interval in the Reconstruction}
\label{Sec:Depth Interval}
The axial interval $\delta_z'$ between the reconstructed cross-section images ($x$-$y$ images) is determined by $\delta'_z=\frac{\lambda_c^2}{2\Delta'_\lambda}$, where $\lambda_c$ is the center wavelength of the SLD source and $\Delta'_\lambda$ is the width of the reconstructed spectrum. For the experiments described in the main text, the full width at $1/20$ maximum of the spectrum (after the filters) is measured to be $40nm$, which gives $\delta'_z=8\mu m$. Note the difference between the axial interval $\delta'_z$ and the axial resolution $\delta_z$ in Eq. (3) in the main text.  $\delta_z'$ represents the sampling interval in the depth dimension, which is calculated from the fast Fourier transform (FFT) of the discrete spectral dimension, whereas $\delta_z$ represents the theoretical physical depth resolution which is calculated from the continuous Fourier transform of the source spectrum (Eq. (3) assumes a Gaussian spectrum). 
Also note $\Delta'_\lambda$ is the sampling range in the spectral dimension while $\Delta_\lambda$ in Eq. (3) denotes the full width at half maximum of the source spectrum.

\section{Details of the CS Algorithm}
\label{Sec:Algorithm details}
Different from other snapshot CS imaging systems, our system translates the depth information into spectral domain. Therefore, one question is whether should the priors be imposed on the datacube of $(x,y,z)$ or $(x,y,\lambda)$? Though it might be more intuitive  to impose the prior on the $(x,y,z)$, this will be a complex-number datacube, \ie, including real parts and imaginary parts.  Handling complex number requires more computational resources as well as running time in addition to the Fourier transforms in each iteration. Therefore, by a large number of experiments, we have found that imposing the priors, both total variation (TV) and wavelet sparsity, on the all real number $(x,y,\lambda)$ datacube gives better results within a shorter running time.

We propose to use a joint constraint with both total variation (TV) and sparsity priors, \ie, $R(\xv) = {\rm TV}(\xv) + \Thetamat(\xv)$, where $\Thetamat(\xv)$ denotes the sparse prior in wavelet domain. We thus solve the following problem
	\begin{equation}
	\hat{\xv} = \argmin_{\xv}\frac{1}{2}\|\yv-\Phimat \xv\|_2^2+\lambda {\rm TV}(\xv) + \rho \Thetamat(\xv),
	\end{equation}
	where $(\lambda, \rho)$ are used to balance the two priors.
For different signals of interest, we can change these parameters to achieve best results.

We solve it via the alternating direction multiplication minimization (ADMM) framework~\cite{Boyd11ADMM} by formulating it as
	\begin{eqnarray}
	(\hat\xv, \hat\zv) &=& \argmin_{\xv,\zv}  \frac{1}{2}\|\yv-\Phimat \xv\|_2^2+\lambda {\rm TV}(\xv) + \rho \Thetamat(\zv), \\
	&{\rm s.t.}& \xv = \zv.
	\end{eqnarray}
	This is solved iteratively by the following sub-problems
	\begin{eqnarray}
	\hat\xv^{k+1} &=& \argmin_{\xv}\frac{1}{2}\|\yv-\Phimat \xv^k\|_2^2+\lambda {\rm TV}(\xv^k) \nonumber\\
	&&+ \frac{\tau}{2}\|\xv^k - \zv^k - \uv^k\|_2^2, \label{Eq:x_k+1}\\
	\hat{\zv}^{k+1} &=& \argmin_{\zv} \frac{\tau}{2}\|\xv^{k+1} - \zv^k - \uv^k\|_2^2 + \rho \Thetamat(\zv), \label{Eq:z_k+1}\\
	\uv^{k+1} &=& \uv^{k} + (\zv^{k+1} -\xv^{k+1}), \label{Eq:u_k+1}
	\end{eqnarray}
	where the superscript $^k$ denotes the iteration number and $\tau$ is a balancing parameter.
	
For the problem in \eqref{Eq:z_k+1}, it is a wavelet based denoising problem, and hereby we employ the soft thresholding algorithm to solve it. Specifically, let $\Tmat$ being the 3D wavelet basis, which can be implemented as $\Tmat = \Tmat_x \otimes \Tmat_y \otimes \Tmat_{\lambda}$, where $\otimes$ denotes the Kronecker product and $\Tmat_x$, $\Tmat_y$ and $\Tmat_\lambda$ denotes the basis along horizontal, vertical and spectral axis, respectively. 
We now have 
\begin{equation}
    \cv = \Tmat\inv \zv
\end{equation}
being the coefficients of the signal $\zv$ in the wavelet domain and impose them to be sparse. \eqref{Eq:z_k+1} can now be solved via
\begin{equation}
    \hat{\cv}^{k+1} = \argmin_\cv \frac{\tau}{2}\|\xv^{k+1} - \uv^k- \zv^k \|_2^2 + \rho \|\cv\|_1, ~~ s.t.~~ \zv = \Tmat \cv,
\end{equation}
where we have used the $\ell_1$-norm to impose the sparsity as well as a tractable answer.
This admits the following solution~\cite{Daubechies04_soft}:
\begin{equation}
\hat{c}^{k+1} = sign(\xv^{k+1} - \uv^k) \odot max(|\xv^{k+1} - \uv^k|-T,0), \label{Eq:c_k}
\end{equation}
where $\odot$ denotes the element-wise product and both the $sign$ (get the sign of the ensured entry) and $|~|$ are taken element-wise. $T>0$ is a thresholding parameter depending on $\tau$ and $\rho$.
After $\hat{\cv}^{k+1}$ is solved, we can get
\begin{equation}
 \hat{\zv}^{k+1} = \Tmat \hat{\cv}^{k+1}.  \label{Eq:z_k}
\end{equation}

As the updating for \eqref{Eq:u_k+1} is straightforward, the left problem is the $\xv$ problem in \eqref{Eq:x_k+1} and we employ ADMM again to solve it by introducing additional variables $\{\pv, \vv\}$
	\begin{eqnarray}
	\hat\xv^{k+1} &=& \argmin_{\xv}\frac{1}{2}\|\yv-\Phimat \xv^k\|_2^2+\frac{\eta}{2}\|\xv^k - (\pv^k + \vv^k)\|_2^2 \nonumber\\
	&&+ \frac{\tau}{2}\|\xv^k - (\zv^k + \uv^k)\|_2^2, \label{Eq:vx_k+1}\\
	\hat\pv^{k+1} &=& \argmin_{\pv} \frac{\eta}{2}\|\pv^k -(\xv^k - \vv^k)\|_2^2 + \lambda {\rm TV}(\pv^k), \label{Eq:pk+1}\\
	\vv^{k+1} &=& \vv^k + (\pv^{k+1} - \xv^{k+1}). \label{Eq:vk+1}
	\end{eqnarray}

\begin{algorithm}[H]
	\label{algo:PnPTVFFD}
		\caption{ADMM-TV-Wavelet for reconstruction}
		\begin{algorithmic}[1]
			\Require$\Phimat$, $\yv$.
			\State Initial $\vv^{0},\pv^0,\uv^0,\zv^0,\eta,\lambda,\rho, \tau$.
			\For {$k=1$ to MaxIter}
			\State Update $\xv$ by \eqref{Eq:x_k+1_sol};
			\State Update $\pv$ by TV denoising algorithm;
			\State Update $\vv$ by \eqref{Eq:vk+1};
			\State Update $\cv$ by \eqref{Eq:c_k};
			\State Update $\zv$ by \eqref{Eq:z_k};
			\State Update $\uv$ by \eqref{Eq:u_k+1};
			\EndFor
			\State $\textbf{Output:}$ Reconstructed signal $\hat \xv$.
		\end{algorithmic}	
\end{algorithm}
	
Eq~\eqref{Eq:vx_k+1} is a quadratic form and has a closed-form solution due to the image-space coding regime of our approach. Below we give a detailed derivation.

Since $\Phimat\Phimat\ts$ is a diagonal matrix, this can be solved very efficiently via
	\begin{equation}
	[\Phimat\ts\Phimat + (\eta+\tau)\Imat]\xv^{k+1} = \Phimat\ts\yv + \eta(\pv^k + \vv^k) + \tau (\zv^k + \uv^k).
	\end{equation}
	Using 
		\begin{align}
 	&[\Phimat\ts\Phimat + (\eta+\tau)\Imat]\inv = (\eta+\tau)\inv \nonumber\\
	&-(\eta+\tau)\inv \Phimat\ts(\Imat + \Phimat(\eta+\tau)\inv\Phimat\ts)\inv\Phimat(\eta+\tau)\inv, \\
	&\Phimat\Phimat\ts = {\rm diag}\{\psi_1, \dots, \psi_N\},\quad N =N_x (N_y+N_\lambda -1),\\
	& \psi_n = \sum_{j=1}^{2N_t} c^2_{j,n}, \forall n=1,\dots, N, \quad \cv_j = {\rm vec}(\Mmat^*_j),\\
	&\tilde{\zv} \stackrel{\rm def}{=}   \eta(\pv^k + \vv^k) + \tau (\zv^k + \uv^k),
	\end{align}
 	we have
	\begin{align}
 	\xv^{k+1} 
 	&=\frac{\Phimat\ts\yv + \tilde{\zv}}{\eta+\tau}- \frac{\Phimat\ts(\Imat + \Phimat(\eta+\tau)\inv\Phimat\ts)\inv \Phimat\Phimat\ts\yv}{(\eta+\tau)^2} \nonumber\\
 	&\qquad- \frac{\Phimat\ts(\Imat + \Phimat(\eta+\tau)\inv\Phimat\ts)\inv \Phimat\tilde{\zv}}{(\eta+\tau)^2}. 
 	\end{align}
 	Since
 	\begin{align}
	&(\Imat + \Phimat(\eta+\tau)\inv\Phimat\ts)\inv 
	= {\rm diag}\left\{\frac{\eta+\tau}{\eta+\tau+\psi_1},\dots, \frac{\eta+\tau}{\eta+\tau+\psi_N}\right\}, \\
	&(\Imat + \Phimat(\eta+\tau)\inv\Phimat\ts)\inv \Phimat\Phimat\ts \nonumber\\
	&=  {\rm diag}\left\{\frac{(\eta+\tau)\psi_1}{\eta+\tau+\psi_1},\dots, \frac{(\eta+\tau)\psi_N}{\eta+\tau+\psi_N}\right\}.
	\end{align}
	We have
	\begin{align}
	\xv^{k+1} 
	=& \frac{\Phimat\ts\yv + \tilde{\zv}}{\eta+\tau} - \Phimat\ts \frac{\left\{\frac{\psi_1 y_1}{\eta+\tau+\psi_1},\dots, \frac{\psi_N y_N}{\eta+\tau+\psi_N}\right\}}{\eta+\tau}\nonumber\\
	&-\Phimat\ts\frac{\left\{\frac{[\Phimat\tilde{\zv}]_1}{\eta+\tau+\psi_1},\dots, \frac{[\Phimat\tilde{\zv}]_N}{\eta+\tau+\psi_N}\right\}}{\eta+\tau} \\
	 =& \frac{\tilde{\zv}}{\eta+\tau} + \Phimat\ts \left[\frac{y_1 - \frac{[\Phimat\tilde{\zv}]_1}{\eta+\tau}}{\eta+\tau+\psi_1},\dots, \frac{y_N - \frac{[\Phimat\tilde{\zv}]_N}{\eta+\tau}}{\eta+\tau+\psi_N}\right]\ts, \label{Eq:x_k+1_sol}
	\end{align}
	where $[\Phimat\tilde{\zv}]_i$ denotes the $i$-th element of the vector in $[~]$. 
	This can be solved element-wise in one shot. \eqref{Eq:pk+1} is a TV-denoising algorithm and can be solved using for example~\cite{Yuan16ICIP_GAP,Bioucas-Dias07_twist}. 
	The entire algorithm, dubbed ADMM-TV-Wavelet, is exhibited in Algorithm 1. 

One left question is to fine tune the parameters, \ie, $\lambda$ and $\rho$, to achieve optimal results.
The main challenge of using this iteration based optimization algorithm, as we mentioned in the main paper, is the long running time.

These challenges, hopefully will be overcome by the advances of deep learning, which is described in Section~\ref{Sec:DL_al}.

\bibliographystyle{IEEEtran}

\end{document}